\documentclass[lettersize,journal]{IEEEtran}
\usepackage{amsmath,amsfonts}
\usepackage{xcolor}
\usepackage{algorithm}
\usepackage{array}
\usepackage{textcomp}
\usepackage{stfloats}
\usepackage{url}
\usepackage{verbatim}
\usepackage{graphicx}
\usepackage{cite}
\usepackage{color}
\usepackage{algorithmicx}   
\usepackage{algpseudocode}  
\usepackage{amsmath}        
\usepackage{color}
\usepackage[caption=false]{subfig}
\usepackage{graphicx}
\usepackage{lineno}
\begin{document}

\title{A Noise Constrained Diffusion (NC-Diffusion) Framework for High Fidelity Image Compression}

\author{Zhenyu Du, Yanbo Gao, Shuai Li,~\IEEEmembership{Senior Member,~IEEE}, Yiyang Li, \\Hui Yuan,~\IEEEmembership{Senior Member,~IEEE}, Mao Ye,~\IEEEmembership{Senior Member,~IEEE}
	\thanks{Z. Du, Y. Gao, S. Li, Y. Li and H. Yuan are with Shandong University, Jinan, China. E-mail:  shuaili@sdu.edu.cn. \par  M. Ye is with University of Electronic Science and Technology of China, Sichuan, China. E-mail: cvlab.uestc@gmail.com   
	
}
}


\maketitle

\begin{abstract}

With the great success of diffusion models in image generation, diffusion-based image compression is attracting increasing interests. However, due to the random noise introduced in the diffusion learning, they usually produce reconstructions with deviation from the original images, leading to suboptimal compression results. To address this problem, in this paper, we propose a \textbf{N}oise \textbf{C}onstrained \textbf{Diffusion} (\textbf{NC-Diffusion}) framework for high fidelity image compression. Unlike existing diffusion-based compression methods that add random Gaussian noise and direct the noise into the image space, the proposed NC-Diffusion formulates the quantization noise originally added in the learned image compression as the noise in the forward process of diffusion. Then a noise constrained diffusion process is constructed from the ground-truth image to the initial compression result generated with quantization noise. The NC-Diffusion overcomes the problem of noise mismatch between compression and diffusion, significantly improving the inference efficiency. In addition, an adaptive frequency-domain filtering module is developed to enhance the skip connections in the U-Net based diffusion architecture, in order to enhance high-frequency details. Moreover, a zero-shot sample-guided enhancement method is designed to further improve the fidelity of the image. Experiments on multiple benchmark datasets demonstrate that our method can achieve the best performance compared with existing methods. 
\end{abstract}
\begin{IEEEkeywords}
image compression, diffusion models, quantization noise, deep learning.
\end{IEEEkeywords}

\section{Introduction}
\IEEEPARstart{D}{ata} compression is a fundamental task in data transmission and processing, by compressing data to a lower bitrates without losing its main information. Currently, learned image/video compression methods \cite{guo2020variable,balle2016end,wang2020ensemble,balle2018variational,lin2022dmvc,zhang2025learning,pan2021tsan}, using deep neural networks as encoder and decoder, have gained increasing interests and achieved superior performance over traditional image/video compression methods \cite{wallace1991jpeg,christopoulos2000jpeg2000,sullivan2012overview,bross2021overview,bellard2015bpg,wang2011ssim}. Existing methods mainly consist of analysis and synthesis transform, quantization and lossless entropy coding, optimized with rate-distortion (RD) cost. Due to the quantization process, noise is introduced in the encoding and decoding process, and with the Mean Squared Error (MSE) based loss function in the RD cost, these learned image compression methods usually lead to relatively blurry and ambiguous results.

\begin{figure}[h]
    \centering
    \subfloat[Existing diffusion-based image compression methods start inference from pure noise and introduce variance perturbations $\epsilon$ during the process, where the condition $c$ can be an initial compressed image or a image feature.\label{fig1new:sub1}]{
        \includegraphics[width=0.5\textwidth]{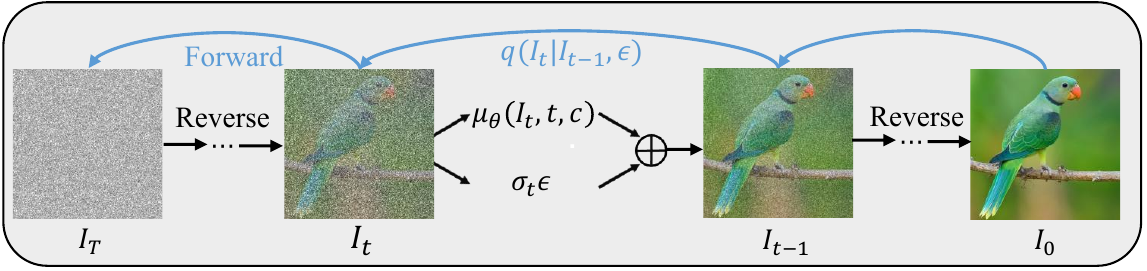}
    }
    
    \vspace{0.5em} 
    
    \subfloat[The proposed method replaces Gaussian noise with quantization-induced noise $\epsilon_n$, enabling direct quantization noise elimination starting from the initially compressed image through a deterministic sampling process $I_{t-1} = f_{\theta}(I_t, t, I_T)$ without introducing randomness.\label{fig1new:sub2}]{
        \includegraphics[width=0.5\textwidth]{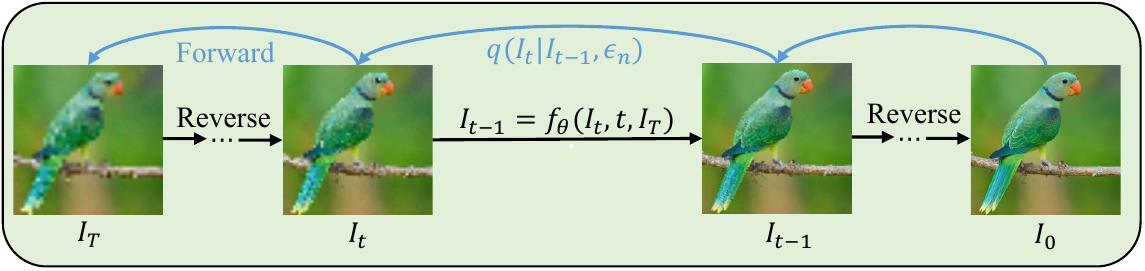}
    }
    \caption{A comparison between existing diffusion-based image compression methods and the proposed method in this paper.}
    \label{fig1}
\end{figure}
\begin{figure*}[htbp]
    \centering
    \subfloat[ Image compression diagram, where the quantization injects the noise $\epsilon _q$ and the decoder $D(\cdot)$ aims to remove such noise.\label{fig1:sub1}]{
        \includegraphics[width=1\textwidth]{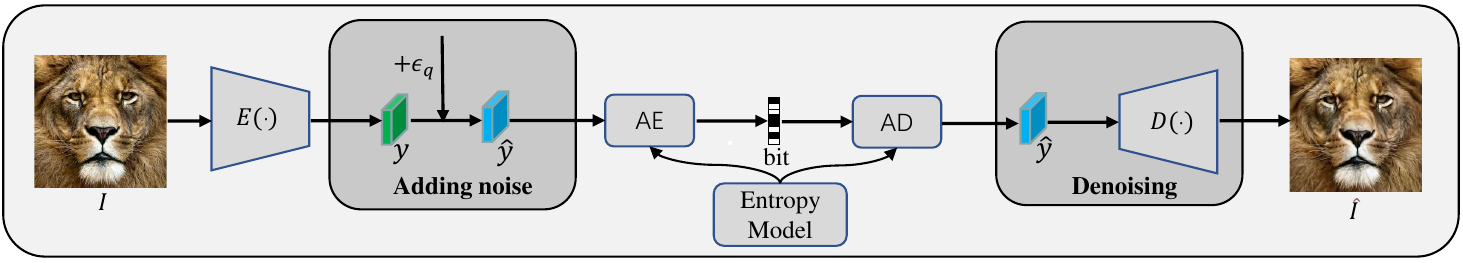}
    }
    \hfill
    \subfloat[ Conditional diffusion model diagram, where the random Gaussian noise is gradually added to the image and then gradually denoised in the reverse process guided by the condition $c$.\label{fig1:sub2}]{
        \includegraphics[width=1\textwidth]{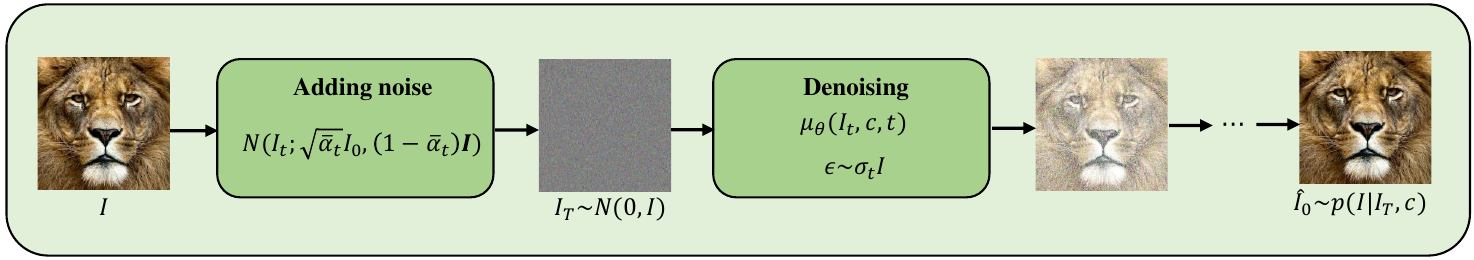}
    }
    \caption{Illustration of the image compression and the diffusion model, both of which can be viewed as a process of noise addition and denoising.}
    \label{fig1:main}
\end{figure*}
To improve the perceptual quality of the decoding results, there have been some studies using generative models in image compression. Especially with the rapid development of diffusion-based methods, denoising diffusion probabilistic models (DDPMs) \cite{ho2020denoising,song2021denoising} have also been investigated for image compression tasks. Existing diffusion-based image compression methods can be categorized into two approaches based on the domain performing the diffusion. The first approach \cite{yang2024lossy,ghouse2023residual,hoogeboom2023high} performs the diffusion in the image domain based on conditions such as initial reconstruction result or image features. The second approach \cite{lei2023text+,pan2022extreme,li2024misc,li2024towards,xue2025one,careil2023towards} operates in the latent domain similar as the Stable Diffusion \cite{rombach2022high}. It uses semantic information such as edge map or texture description as condition to produce results that are semantically consistent. This category of methods \cite{lei2023text+,pan2022extreme,li2024misc,li2024towards,xue2025one,careil2023towards} usually target at low bitrates image compression and cannot perform well in terms of objective reconstruction quality. Both approaches start the generation process from random Gaussian noise, as shown in Fig. \ref{fig1new:sub1}. This introduces inherent uncertainty and additional noise throughout the diffusion process. It is good for image generation task by introducing diversity into the generation process. However, they usually cannot provide faithful reconstructions of the original to-be-encoded images due to their random generation nature. This can be catastrophic for the compression task in some usage scenarios. Currently, there is no thorough investigation on solving the random property in diffusion models for realistic image compression in order to fully explore its generation capability yet. 

In this paper, we first investigate the relationship between diffusion model and image compression, where the diffusion model and image compression are bridged together through noise addition and noise removal, as shown in Fig. \ref{fig1:sub1} and Fig. \ref{fig1:sub2}. However, the diffusion model starts the reconstruction from Gaussian random noise, whereas the noise introduced in the image compression task is quantization noise at the latent feature before entropy and decoder-specific noise at the reconstructed image level. Consequently, there is a noise distribution mismatch problem, as shown in Fig. \ref{fig2:main}, which leads to additional noise in the reconstruction. Note that the latent feature before entropy in image compression is different from that in latent diffusion model such as Stable Diffusion \cite{rombach2022high}, where spatial patterns are optimized to be removed in image compression and thus difficult to directly remove the additive quantization noise.

To solve the above noise distribution mismatch induced random reconstruction problem, a NC-Diffusion framework is proposed for image compression in this paper. The quantization and the decoding transform is regarded as the noise addition process, where the quantization noise is equivalent to the added random noise in diffusion but constrained. In other words, the proposed NC-Diffusion framework is designed to diffuse the ground-truth image into the quantized noisy image, and thus in the inference stage, it can start the inference directly from the initial decoded image to enhance the faithfulness of reconstruction. Fig. \ref{fig1new:sub2} further illustrates the difference between the proposed method and the existing diffusion based methods. By modeling on the quantization noise in the forward diffusion process, it avoids introducing further uncertainty by adding random Gaussian noise as in the existing methods. In addition, to further enhance the high-frequency details, a frequency-domain adaptive filtering module is developed to enhance the skip connections in the U-Net based diffusion network. Moreover, a zero-shot sample-guided enhancement method is used to further improve the fidelity of the reconstructed image.

Our contributions can be summarized as follows:
\begin{itemize}
\item[$\bullet$] We propose a NC-Diffusion framework for high fidelity image compression. The method overcomes the problem of noise mismatch when applying diffusion to image compression, thus improving the perceptual performance without introducing extra random noise. 
\item[$\bullet$] We propose a plug-and-play frequency-domain adaptive filtering module. It can be applied to any diffusion model of U-Net architecture to enhance the skip connections in the frequency-domain. In addition, a zero-shot sample-guided enhancement is used at test to further improve perceptual quality.
\item[$\bullet$] Extensive experiments demonstrate that the proposed method achieves better performance than the existing methods in terms of rate-distortion and rate-perception.
\end{itemize}

\section{Related Work}
This section briefly describes the related work on diffusion-based generative image compression, as well as the methods using diffusion models for other low-level vision tasks.

\subsection{Generative Image Compression}

In addition to employing Convolutional Neural Network architectures for image compression \cite{li2024high,ge2025rethinking,zhang2024machine} and video coding \cite{jia2023mpai,sheng2024spatial,jin2021deep}, methods based on generative models have been widely investigated in the last few years. It takes advantage of the generation capability to enhance the perceptual quality of the image compression results. In \cite{mentzer2020high,muckley2023improving}, the adversarial loss proposed in the generative adversarial network \cite{goodfellow2014generative} was first incorporated into the learnable image compression to achieve the rate-distortion-perceptual trade-off. With the diffusion model achieving impressive performance on image generation, diffusion-based generative image compression \cite{theis2022lossy,yang2024lossy,khoshkhahtinat2024laplacian,ghouse2023residual,hoogeboom2023high} has been actively studied. Theis {\em et al.} \cite{theis2022lossy} proposed to transfer a noisy image added with Gaussian noise via reverse channel encoding and use an unconditional diffusion model to generate the image based on the received noisy image. In contrary, a conditional diffusion model based image compression was proposed in \cite{yang2024lossy}, which performs inference from random noise and uses the feature encoded by a variational autoencoder (VAE) \cite{kingma2013auto} as condition. Atefeh {\em et al.} \cite{khoshkhahtinat2024laplacian} further proposed to consider the relative importance of different frequency components and diffuse each frequency component at distinct rates, thus leading to a coarse-to-fine generation of decoded images. Ghouse {\em et al.} \cite{ghouse2023residual} and Hoogeboom {\em et al.} \cite{hoogeboom2023high} used a diffusion model at the decoder-side to enhance the compression results from existing learned image compressor. Compared with \cite{yang2024lossy,khoshkhahtinat2024laplacian}, it uses an initial compression result as condition for the diffusion model to generate the final image from Gaussian noise. However, the above models, diffusing from random Gaussian noise, introduces randomness into the compression results and usually takes a large number of inference steps to propagate the noise to the desired results. 

\begin{figure}[t]
    \centering
    \subfloat[Random Gaussian noise.\label{fig2:sub1}]{
        \includegraphics[width=0.48\textwidth]{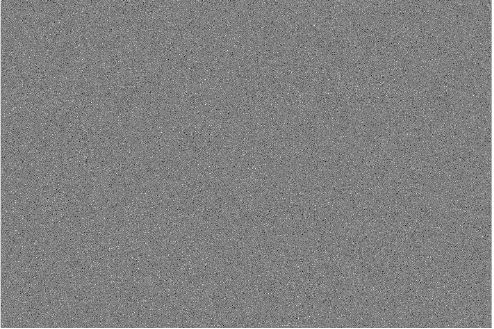}
    }
    
    \vspace{0.5em} 
    
    \subfloat[Noise generated on $kodim23$.\label{fig2:sub2}]{
        \includegraphics[width=0.48\textwidth]{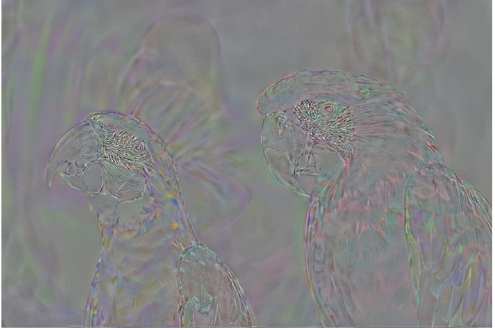}
    }
    \caption{Comparison of the normalized noises generated in the image compression and added in the diffusion model. The pattern around edges in the noise generated in image compression and the random noise clearly illustrates the noise mismatch problem if directly using diffusion model for image compression.}
    \label{fig2:main}
\end{figure}
There are also some works \cite{lei2023text+,pan2022extreme,li2024misc,li2024towards,xue2025one,careil2023towards} aiming to enhance the perceptual performance of image compression at a very low bitrates by utilizing the generative ability of diffusion models. Lei {\em et al.} \cite{lei2023text+} proposed to compress only the prompt inversion text and edge detection map at very low bitrates and then use ControlNet to generate the image at the decoder-side. Similarly in \cite{pan2022extreme,li2024misc}, a textual description obtained based on a large language model, an initial compressed image and semantic map at a very low bitrates is used as condition for diffusion. Li {\em et al.} \cite{li2024towards} proposed a two-stage model with the first stage to compress images with a VAE-based codec and then uses the decoded content feature to adjust the pretrained Stable Diffusion model with the ControlNet architecture. Xue {\em et al.} \cite{xue2025one} utilized the hyper-prior information as a condition to guide the diffusion in the latent feature domain and introduced a semantic distillation mechanism to enhance the semantic capability of the hyper-prior from  a pretrained generative tokenizer. Careil  {\em et al.} \cite{careil2023towards} utilized a vector-quantized image representation along with a global image description in order to reduce the bitrate of the condition. Such low bitrates models generally focus on generating a perceptually good image with similar semantic information over the consistency to the original image.

\subsection{Diffusion Models in Low-Level Vision}

Diffusion models are widely used in a wide variety of low-level vision tasks in addition to image compression, such as image super-resolution \cite{saharia2022image,li2022srdiff,shang2024resdiff,rombach2022high,yue2024resshift,wang2024sinsr}, restoration \cite{garber2024image,liu2024diff,liu2024residual,lin2024diffbir,kawar2022denoising,ren2024MoE-DiffIR}, low-light enhancement \cite{jiang2023low,he2024zero,zhou2023pyramid,jiang2024lightendiffusion}, deblurring \cite{whang2022deblurring,ren2022image,xia2023diffir,delbracio2023inversion} and other tasks. In \cite{saharia2022image}, the diffusion model is used for image super-resolution, where the noise is used as input to the diffusion model with the low-resolution image  as condition. Instead of directly generating a high-resolution image, Li {\em et al.} \cite{li2022srdiff} used diffusion to generate the residuals between low-resolution and high-resolution images. Shang {\em et al.} \cite{shang2024resdiff} further used a CNN-based image super-resolution network to extract the features of a low-resolution image as condition. In addition to directly performing diffusion on the image level, Rombach {\em et al.} \cite{rombach2022high} performed the diffusion process in the latent space to reduce the computational complexity, and the low-resolution image features were injected into the diffusion model as condition in a cross-attention manner. There are also some methods \cite{yue2024resshift,wang2024sinsr} that use diffusion models to construct residual shifting between low-resolution and high-resolution images, where the residual is progressively added together with the Gaussian noise to the high-resolution image according to noise schedule. Whang {\em et al.} \cite{whang2022deblurring} used the diffusion model for image deblurring, by using a deterministic predictor as condition for the diffusion model. Ren {\em et al.} \cite{ren2024MoE-DiffIR} proposed to use mixture-of-experts (MOE) to combine different prompts and a visual-to-text adapter was also used to explore cross-modality generation prior in Stable Diffusion. Jiang {\em et al.} \cite{jiang2023low} proposed to use diffusion model on the low-frequency component via wavelet decomposition for low-light image enhancement,  and the enhanced low-frequency component is combined with the restored high-frequency components to achieve high fidelity performance. The above diffusion-based methods on various low-level visual tasks illustrate that the diffusion model has great potential for image refinement tasks in addition to generative tasks. However, a deterministic generation, consistent to the original image without deviation due to the random Gaussian noise added in the diffusion learning, still needs further investigation.

\section{Motivation}
\begin{figure*}[t]
  \centering
    \includegraphics[width=1\textwidth]{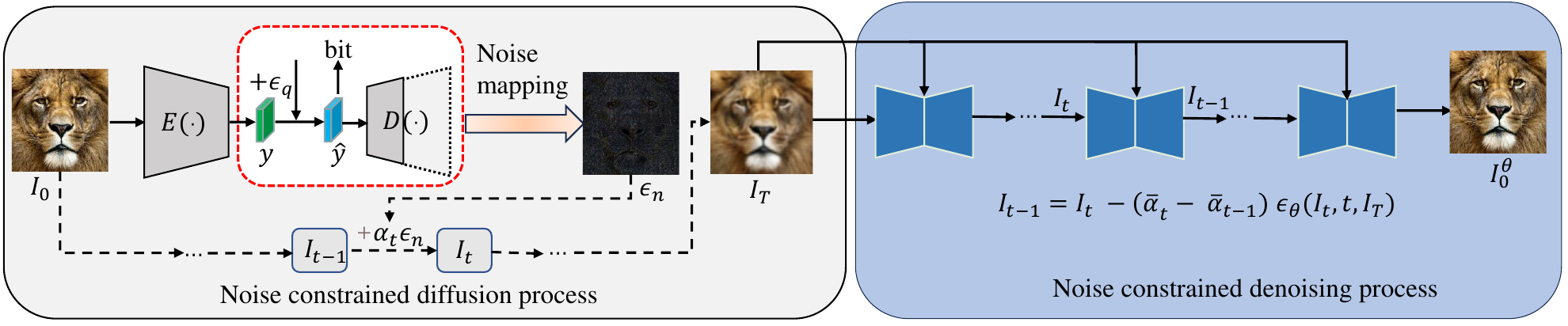}
    \caption{Overview of the proposed NC-Diffusion framework for image compression. At initial compression, a quantization noise $\epsilon_{q}$ is introduced to produce a noisy feature $\hat{y}$. A noise constrained forward process is constructed for the noise $\epsilon_{n}$ existing between the ground-truth image $I_0$ and the initial compression result $I_T$, where $\epsilon_{n}$ is added according to a noise schedule $\alpha_{t}$. A NC-Diffusion $\epsilon_{\theta}(I_{t},t,I_{T})$ is trained to reverse the forward process in order to achieve denoising directly from the initial compression result $I_T$ to eliminate the noise $\epsilon_{n}$. Note that the dotted line in $D(\cdot)$ indicates that NC-Diffusion can be placed anywhere in the decoder for denoising.}
    \label{fig3:framework}
\end{figure*}
Image compression is generally composed of encoding analysis transform, quantization, entropy coding, and decoding synthesis transform as shown in Fig. \ref{fig1:sub1}. It can be considered as a noise addition and noise removal process, where the quantization injects the noise and the decoding aims to remove such noise. On the other hand, the diffusion model consists a forward process, to propagate Gaussian random noise to an image, and a reverse process, to remove noise for inference, which are iteratively performed as shown in Fig. \ref{fig1:sub2}. From the noise addition and removal perspective, image compression and diffusion model work similarly, where noise is added in the encoding/forward process and the model learns to remove noise in the decoding/reverse process. From this perspective, the diffusion model can be directly inserted to the image compression decoder, where the quantization noise is regarded as the added noise in the diffusion forward process.

However, the noise distributions for image compression and diffusion model are different and simply equating them and replacing each other cannot lead to a faithful reconstruction of the original to-be-encoded image. For image compression, the noise is introduced by quantization (usually uniform noise in training and rounding error in test) and then processed through the decoding transform. While the noise added at the latent feature before entropy coding follows uniform distribution, the noise appeared in the reconstructed image or the mid-level features processed by decoder layers cannot be simply characterized by uniform distribution, but follows an unknown conditional joint distribution determined by the decoding transform and initial quantization noise. Specifically, the existing VAE-based image compression architecture cannot compress regions with rich textures high quality, thus showing relatively larger distortion than regions with simple texture. Therefore, patterns with different magnitudes for different textured regions exist as shown in Fig. \ref{fig2:sub2}. On the other hand, for diffusion model, the noise follows Gaussian distribution and is completely random, as shown in Fig. \ref{fig2:sub1}. While the added Gaussian random noise introduces content diversity in image generation, such random noise produces undesired artifacts in image compression. Essentially when using diffusion model to remove noise in image compression, there is a noise distribution mismatch problem, as shown in Fig. \ref{fig2:main}, resulting in a random noise induced poor reconstruction. This paper investigates the noise distribution mismatch problem between the image compression and the diffusion model, and proposes a NC-Diffusion framework.

\section{Proposed Method}

In this section, we present the proposed NC-Diffusion framework customized for image compression. Fig. \ref{fig3:framework} illustrates the overall process. Firstly, a neural image compressor backbone is utilized to encode and decode an image. The NC-Diffusion is inserted in the decoder to reduce the quantization noise from the initial coded result  and enhance the perceptual quality. It adopts the U-Net architecture as the baseline, and an adaptive frequency-domain filtering module is designed to improve the skip connections, in order to enhance the high-frequency details. At test, a zero-shot sample-guided enhancement method is developed to further enhance the perceptual performance. Note that the NC-Diffusion can be placed after different decoder layers, {\em i.e.}, the quantization noise can be removed at different feature levels. In this paper, it is used to directly process the reconstructed image in order to fully explore the knowledge embedded in the decoder.

\subsection{Noise Constrained Diffusion for Image Compression}
As described in the Motivation, directly applying the diffusion method to neural image compression by adding Gaussian random noise suffers from the noise distribution mismatch problem. In the following, the proposed NC-Diffusion is formulated within the image compression process. 

The neural image compression process can be represented as:
\begin{equation}
  \hat{I}\ =\ D\left\{ Q\left[ E\left( I \right) \right] \right\},
  \label{eq:1}
\end{equation}
where $I$ and $\hat{I}$ represent the original image and the decoded image, respectively. $E$ and $D$ denote the encoder and decoder of the neural image compression network, respectively, and $Q$ denotes the quantization. At training, uniform noise is added to the feature to mimic a differentiable quantization process while at test rounding is used as quantization. With quantization, noise is introduced in the compression process, leading to a lossy reconstruction. 
\begin{algorithm}[t]
    \caption{The NC-Diffusion \textbf{training}}
    \label{algo:1} 
    \begin{algorithmic}[1]
        \While{Not converged}
            \State $I_{0} \sim q(I_{0})$
            \State \emph{\# Forward diffusion process}
            \State $t \sim \operatorname{Uniform}\{1,\cdots,T\}$
            \State $\epsilon_{n} = D\{ Q[ E( I_{0}) ] \} - I_{0} = I_{T} - I_{0}$ 
            \State $I_{0}^{\theta} = I_{0} + \bar{\alpha}_{t}\epsilon_{n} - \bar{\alpha}_{t}\epsilon_{\theta}( I_{0} + \bar{\alpha}_{t}\epsilon_{n}, t, I_{T})$
            \State Perform a single gradient descent step for:
            \State \begin{footnotesize}
                $\begin{aligned}
                &\nabla_{\theta} \left\| \epsilon_{\theta}(I_{0} + \bar{\alpha}_{t}\epsilon_{n}, t, I_{T}) - \epsilon_{n} \right\|_{2}^{2} + \omega \, d_{\text{LPIPS}}(I_{0}, I_0^\theta)\\
                &+ \beta \sum_{i=1}^{K} E\left[ \left\| H_{i} - \widehat{H}_{i} \right\|_{2}^{2} + \left\| V_{i} - \widehat{V}_{i} \right\|_{2}^{2} + \left\| D_{i} - \widehat{D}_{i} \right\|_{2}^{2} \right]
                \end{aligned}$
                \end{footnotesize}
        \EndWhile
    \end{algorithmic}
\end{algorithm}

Considering that diffusion model is learned with noise addition and removal, in this paper, the diffusion model is used in the decoder by taking the quantization and the decoding transform as the noise addition process. Accordingly, the quantization noise after the decoding transform is equivalent to the added random noise in diffusion model. In this way, the diffusion model can be explored to enhance the perceptual quality without introducing additional noise. Specifically, the noise between the input image and the reconstructed image can be obtained by:
\begin{equation}
    \epsilon_n=\hat{I}-I=D\{Q[E(I)]\}-I=f(\epsilon_q),
  \label{eq:2}
\end{equation}
where $\epsilon _q$ represents the quantization noise introduced on the compressed feature $E\left( I \right)$. Assuming the encoder and decoder transforms are ideally trained or adopt coupled transforms such as invertible neural networks, the encoded feature without noise can reconstruct the image losslessly. In such a case, the distortion is only related to the quantization noise and thus can be formulated with a nonlinear function $\epsilon _n=f\left( \epsilon _q \right)$.With $\epsilon _q$ follows a uniform distribution at training, $\epsilon _n$ can be formulated as a conditional probability distribution $p\left( \epsilon _n|\epsilon _q,I,D \right)$ based on the quantization noise $\epsilon _q$, the input image and the decoding transform $D\left( \cdot \right)$. With the complex learned decoding transform, $p\left( \epsilon _n|\epsilon _q,I,D \right)$ cannot be simply approximated with a spatially independent distribution such as the Gaussian distribution. Therefore, in this paper, instead of sampling from a random distribution, actual noise sampled from the noise $\epsilon _n$ in the image domain, obtained by the random quantization noise $\epsilon _q$ added in feature domain and then processed by the fixed nonlinear decoding transform, is used and directly processed in the image domain. In other words, a constrained noise generated from random quantization noise and the decoding transform is used for noise addition in the diffusion model. In this way, the noise addition in the feature domain of image compression is associated with the diffusion process in the image domain of diffusion model.

\begin{algorithm}[t]
    \caption{The NC-Diffusion \textbf{inference}}
    \label{algo:2} 
    \begin{algorithmic}[1]
        \State $I_{0} \sim q(I_{0})$
        \State $I_{T} = D\{Q[E(I_{0})]\}$
        \State \emph{\# Denoising process}
        \For{$t = T \text{ down to } 1$}
            \State $I_{0}^{\theta} = I_{t} - \bar{\alpha}_{t}\epsilon_{\theta}(I_{t}, t, I_{T})$
            \State $L_{\text{CLIP}} = \operatorname{E}[ \|D_{\text{clip}}(I_{0}^{\theta}) - D_{\text{clip}}(I_{T}) \|_{2}^{2} ]$
            \State $I_{t-1} = I_{t} - (\bar{\alpha}_{t} - \bar{\alpha}_{t-1})\epsilon_{\theta}(I_{t}, t, I_{T}) - \lambda\nabla_{I_{0}^{\theta}}L_{\text{CLIP}}$
        \EndFor
        \State \textbf{return} $I_0$
    \end{algorithmic}
\end{algorithm}
With the ground-truth image ($I_0=I$) and {the final noisy image ($I_T=\hat{I}$)} after propagating the noise to $T$ steps, a forward diffusion process can be constructed by adding different scales of $\epsilon _n$ to the ground-truth image $I_0$ according to a variance schedule. In this paper, for simplicity, the mean of the ground-truth image is not scaled in the forward process while only the variance is changed. Thus at step $t$, the NC-Diffusion  forward process can be iteratively obtained via a Markov chain as:
\begin{equation}
  I_t=I_{t-1}+\alpha _t\epsilon _n{\sim}p\left( I_t|I_{t-1},\alpha _t\epsilon _n \right),
  \label{eq:3}
\end{equation}

where $\alpha _t$ represents the coefficient in the variance schedule. Note that $\epsilon _n$ is a zero-mean distribution since the neural compression codec is initially optimized with MSE, and thus the mean of the distribution is not changed in the forward process. The marginal probability of $I_t$ follows a conditional distribution similarly as $\epsilon_n$ but with a smaller variance. $\epsilon _n$ can be viewed as sampling from distribution $p\left( \epsilon _n|\epsilon _q,I,D \right)$ based on the random quantization noise $\epsilon _q$ and the decoder transform. Similarly as in the diffusion model, $I_t$ at arbitrary step can be directly obtained from $I_0$ by: 
\begin{equation}
\begin{aligned}
I_{t}= & I_{t-1}+\alpha_{t}\epsilon_{n} \\
= & I_{t-2}+(\alpha_{t-1}+\alpha_{t})\epsilon_{n} \\
\mathrm{=} & \mathrm{...} \\
= & I_{0}+\bar{\alpha}_{t}\epsilon_{n}{\sim}p(I_{t}|I_{0},\bar{\alpha}_{t}\epsilon_{n})=p\big(I_{t}\big|I_{0},\bar{\alpha}_{t},\epsilon_{q},D\big),
\end{aligned}
\label{eq:4}
\end{equation}
where $\bar{\alpha}_t=\sum_{i=1}^t\alpha_i$, and $\bar{\alpha}_T$ is equal to 1 when the diffusion step $t=T$.

By comparing with the diffusion model, the proposed NC-Diffusion gradually diffuses the reconstruction noise that is determined based on the random quantization noise, input image and the decoding transform, instead of completely random noise. It does not introduce extra randomness that is not already contained in the compression process, and thus overcomes the noise mismatch problem between image compression and diffusion model.

With the NC-Diffusion forward process constructed, the reverse process can be formulated in a similar way as the diffusion model by learning $p\left( I_{t-1}|I_t,I_T \right)$, through a denoising network. This can be achieved by training a noise prediction network $\epsilon _{\theta}\left( I_t,t,I_T \right)$ to predict $\epsilon_n$. Specifically, the general U-Net architecture based noise prediction network is used, and the noisy image $I_t$ is used as the input while the initial neural compression result $I_T$ is used as condition at each step to avoid the diffusion process deviating from the initial result.

By predicting $\epsilon_n$, the predicted image $I_{0}^{\theta}$ at each step can be obtained as: 
\begin{equation}
  I_0^\theta=I_t-\bar{\alpha}_t\epsilon_\theta(I_t,t,I_T), 
  \label{eq:5}
\end{equation}

During the training process, since the marginal distribution of the forward diffusion process is $I_t=I_0+\bar{\alpha}_t\epsilon_n$, and in order to improve the perceptual performance, the following diffusion loss function $L_{diff}$ is used for supervision:
\begin{equation}
  L_{\mathrm{diff}}=E_{I_0,t,\epsilon_n}[\left|\left|\epsilon_\theta(I_t,t,I_T)-\epsilon_n\right|\right|_2^2 + \omega \, d_{\text{LPIPS}}(I_{0}, I_0^\theta)],
  \label{eq:6}
\end{equation}

For the inference, the initial neural compression image $I_T$ is used as the input and progressively denoised by the noise prediction network $\epsilon _{\theta}\left( I_t,t,I_T \right)$. The input at step $t-1$ can be obtained according to Eq. \ref{eq:4} as: 
\begin{equation}
  I_{t-1}= I_0^\theta+ \bar{\alpha}_{t-1}\epsilon_\theta(I_t,t,I_T),
  \label{eq:7}
\end{equation}
where $I_{0}^{\theta}$ and $\epsilon _{\theta}\left( I_t,t,I_T \right)$ are the denoised result and predicted noise at step $t$. With denoised image $I_{0}^{\theta}$ at step $t$ expressed by Eq. \ref{eq:5}, Eq. \ref{eq:7} can be further turned into:
\begin{equation}
  I_{t-1}=I_t-(\bar{\alpha}_t-\bar{\alpha}_{t-1})\epsilon_\theta(I_t,t,I_T),  
  \label{eq:8}
\end{equation}
where the input at step $t-1$ can be directly obtained with the input $I_t$ at step $t$ and the predicted noise, thus forming an iterative denoising process. Accordingly, the initial image $I_T$ can be denoised step by step to approximate the ground-truth image $I_0$.

By designing the above NC-Diffusion specific to the compression task, our model can directly reduce the quantization noise from the initial reconstruction result without introducing extra random noise. Moreover, without inferring from the random Gaussian noise, the proposed NC-Diffusion, directly performing inference from the reconstructed image, can greatly improve the inference efficiency compared with the existing diffusion models. The specific procedure for the NC-Diffusion training and inference is shown in Algorithm \ref{algo:1} and Algorithm \ref{algo:2}, respectively.

\subsection{Adaptive Frequency-domain Filtering}
\begin{figure}[t]
  \centering
    \includegraphics[width=0.48\textwidth]{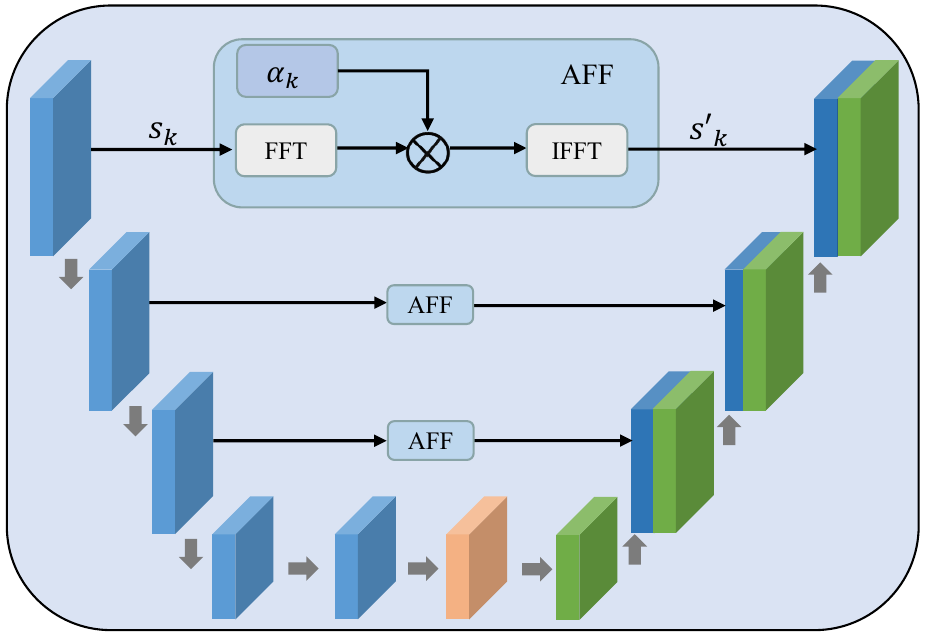}
    \caption{Illustration of the AFF module incorporated in the U-Net architecture to enhance the skip connections.}
    \label{aff}
\end{figure}

It is known that the distortion in the reconstructed image is mostly due to the missing of the high-frequency information, since the original learned codec can restore the low-frequency information relatively well. Therefore, careful design of the high-frequency processing module is desired to further reconstruct the complex high-frequency details lost in the initial image compression network. As illustrated in \cite{si2024freeu}, the backbone of the U-Net architecture in the diffusion model mainly contributes to denoising, and the skip connections are mainly responsible for bringing high-frequency features into the decoder module. An adaptive frequency-domain filtering (AFF) module is developed to enhance the shortcut connection. Fig. \ref{aff} illustrates the AFF module on top of the diffusion backbone. In this paper, a U-Net architecture as in \cite{saharia2022image,shang2024resdiff} is used, where the encoder features are connected to the decoder via an AFF enhanced skip connection.

Specifically, the skip features from the encoder of the U-Net are first transformed into the Fourier frequency-domain and then a spectral modulation is performed to realize high-frequency filtering as:
\begin{equation}
  s_k^{\prime}=f_{IFFT}(\alpha_k\odot f_{FFT}(s_k)),
  \label{eq:9}
\end{equation}
\begin{equation}
  \alpha_k=\begin{cases}1&if\; \; r< r_{\mathrm{th}},\\1+\gamma_k&otherwise.\end{cases}
  \label{eq:10}
\end{equation}
where $s_k$ denotes the skip feature generated by the $k$-th block of the encoder, and $s_k^{\prime}$ denotes the skip feature after high-frequency filtering. $f_{FFT}$ and $f_{IFFT}$ denote the Fourier Transform and its inverse, respectively. $\odot$ denotes the element-wise multiplication. $\alpha_k$ is a learnable mask filled with learned parameter $\gamma_k$ and 1, which serves to adaptively enhance high-frequency features. $r$ represents the frequency radius and $r_{th}$ represents the frequency threshold. $r_{th}$ is set differently for features of different sizes, {\em i.e.}, a larger $r_{th}$ for shallow skip features due to the wider range of frequencies they represent. With the proposed AFF module, the high-frequency components of the skip features are explicitly enhanced without affecting the low-frequency information, thus improving the high-frequency reconstruction.

A high-frequency detail preservation loss function is also added for supervision in the frequency-domain to better reconstruct high-frequency details. The ground-truth image $I_0\in R^{H\times W\times C}$ and the predicted image $I_0^\theta\in R^{H\times W\times C}$ based on the NC-Diffusion are both decomposed into four frequency sub-bands using a multilevel 2D discrete wavelet transformation. The MSE between each high-frequency sub-band is calculated as the high-frequency detail preservation loss function:
\begin{equation}
  \begin{aligned}L_{high}=\sum_{i=1}^KE[||H_i-\widehat{H}_i||_2^2+ &||V_i-\widehat{V}_i||_2^2|\\+||D_i- \widehat{D}_i||_2^2],\end{aligned}
  \label{eq:11}
\end{equation}

where $\{H_i,V_i,D_i\}\text{,}\{\widehat{H}_i,\widehat{V}_i,\widehat{D}_i\}\in R^{\frac H{2^i}\times\frac W{2^i}\times C}$ denote the high-frequency sub-bands of the ground-truth image and predicted image in the horizontal, vertical, and diagonal directions, respectively. $i\in[1,K]$ denote the different levels of wavelet transform.

The total training loss $L_{total}$ for the NC-Diffusion can be obtained by summarizing the general diffusion loss and the high-frequency detail preservation loss as:
\begin{equation}
  L_{total}=L_{diff}+ \beta L_{high},
  \label{eq:12}
\end{equation}
where $\beta$ is a constant factor to balance the two loss terms.
\subsection{Sample-guided Enhancement Method}
To further enhance the perceptual performance of the proposed NC-Diffusion in practical use, a sample-guided perceptual quality enhancement method is proposed. Drawing on the classifier-guidance approach \cite{dhariwal2021diffusion} that utilizes an image classifier trained on a noisy image to guide the diffusion generation process, a perceptual loss between the initial compression result and the enhanced image is constructed to guide the sampling process towards high fidelity.

\begin{figure*}[htbp]
    \centering
    \subfloat[\label{fig4:sub1}]{
        \includegraphics[width=0.48\textwidth]{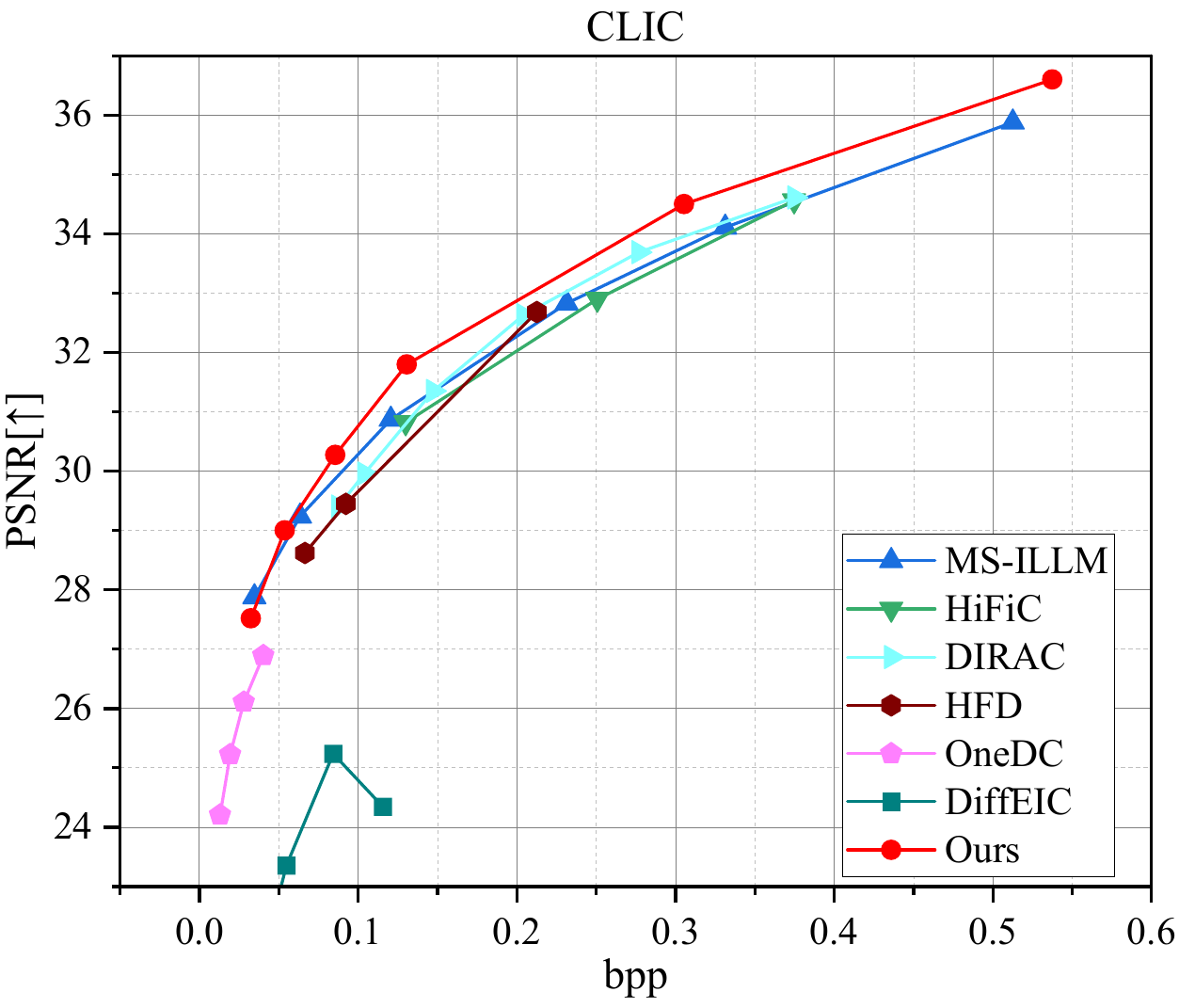}
    }
    \hfill
    \subfloat[\label{fig4:sub2}]{
        \includegraphics[width=0.48\textwidth]{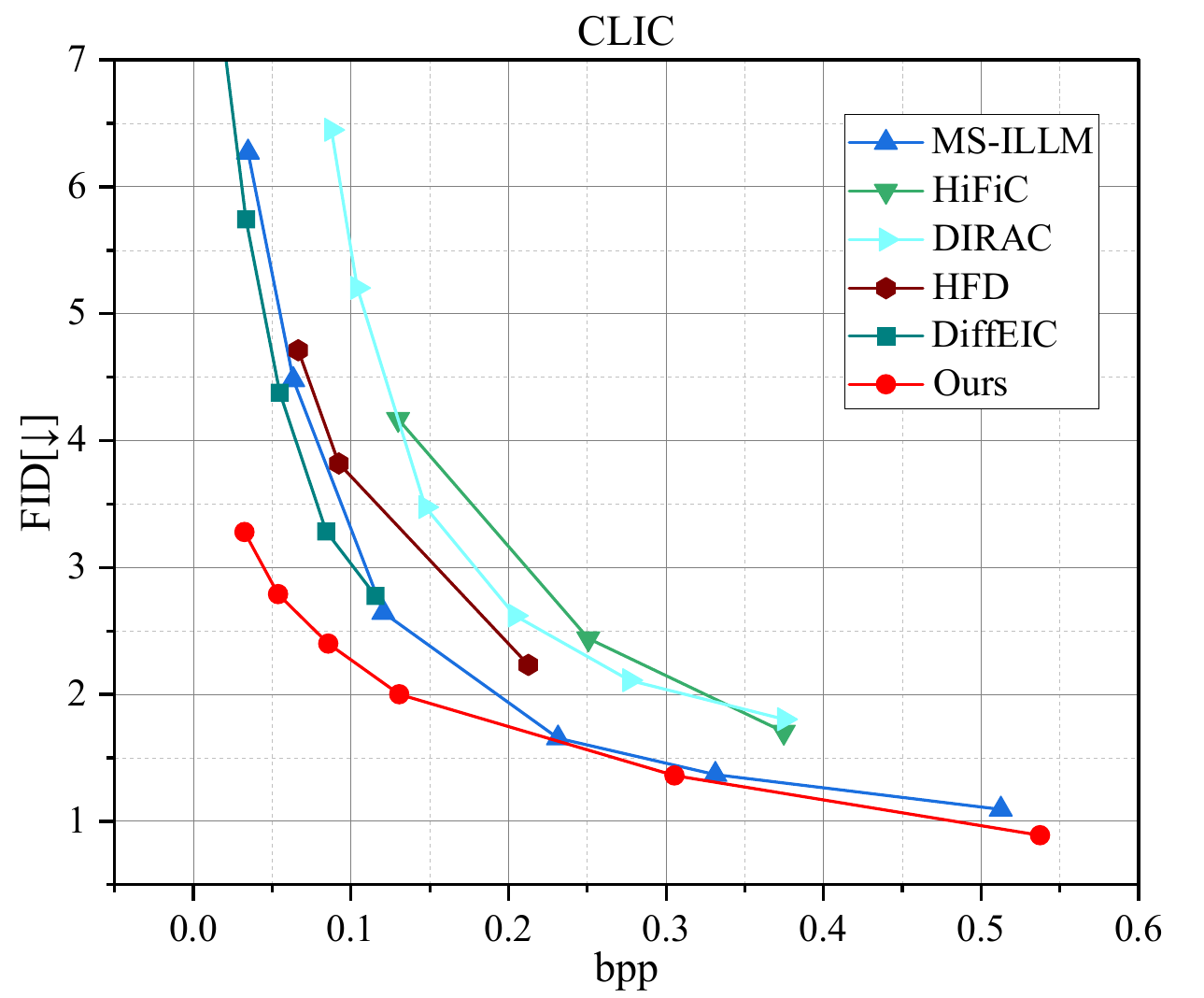}
    }
    \caption{Comparison of our method with existing methods in terms of rate-distortion $[\mathrm{bpp}\downarrow/\mathrm{PSNR}\uparrow]$ and rate-perception$[\mathrm{bpp}\downarrow/\mathrm{FID}\downarrow]$ for the CLIC2020 test dataset.}
    \label{fig4}
\end{figure*}

\begin{figure*}[htbp]
    \centering
    \subfloat[\label{fig5:sub1}]{
        \includegraphics[width=0.31\textwidth]{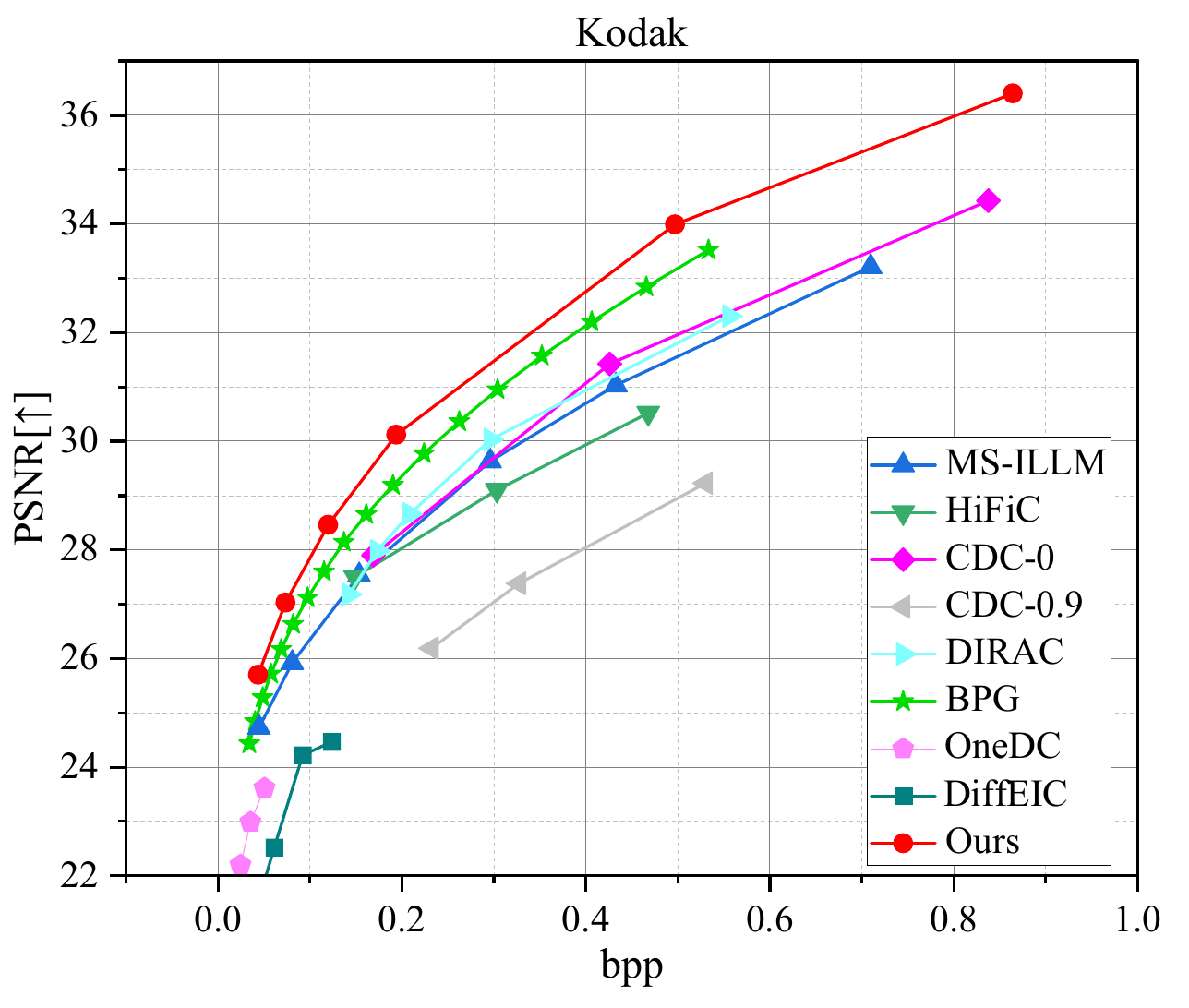}
    }
    \hfill
    \subfloat[\label{fig5:sub2}]{
        \includegraphics[width=0.31\textwidth]{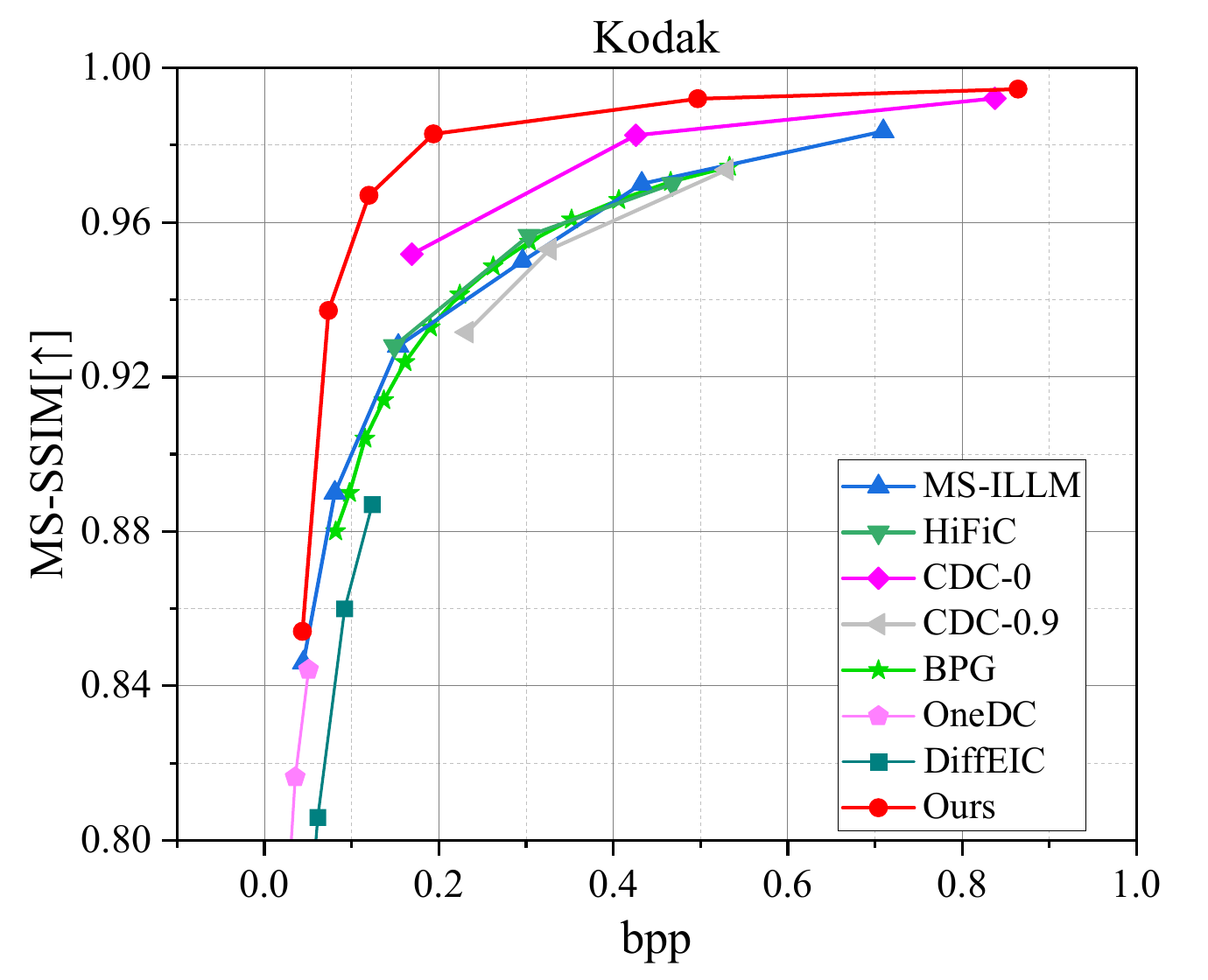}
    }
    \hfill
    \subfloat[\label{fig5:sub3}]{
        \includegraphics[width=0.31\textwidth]{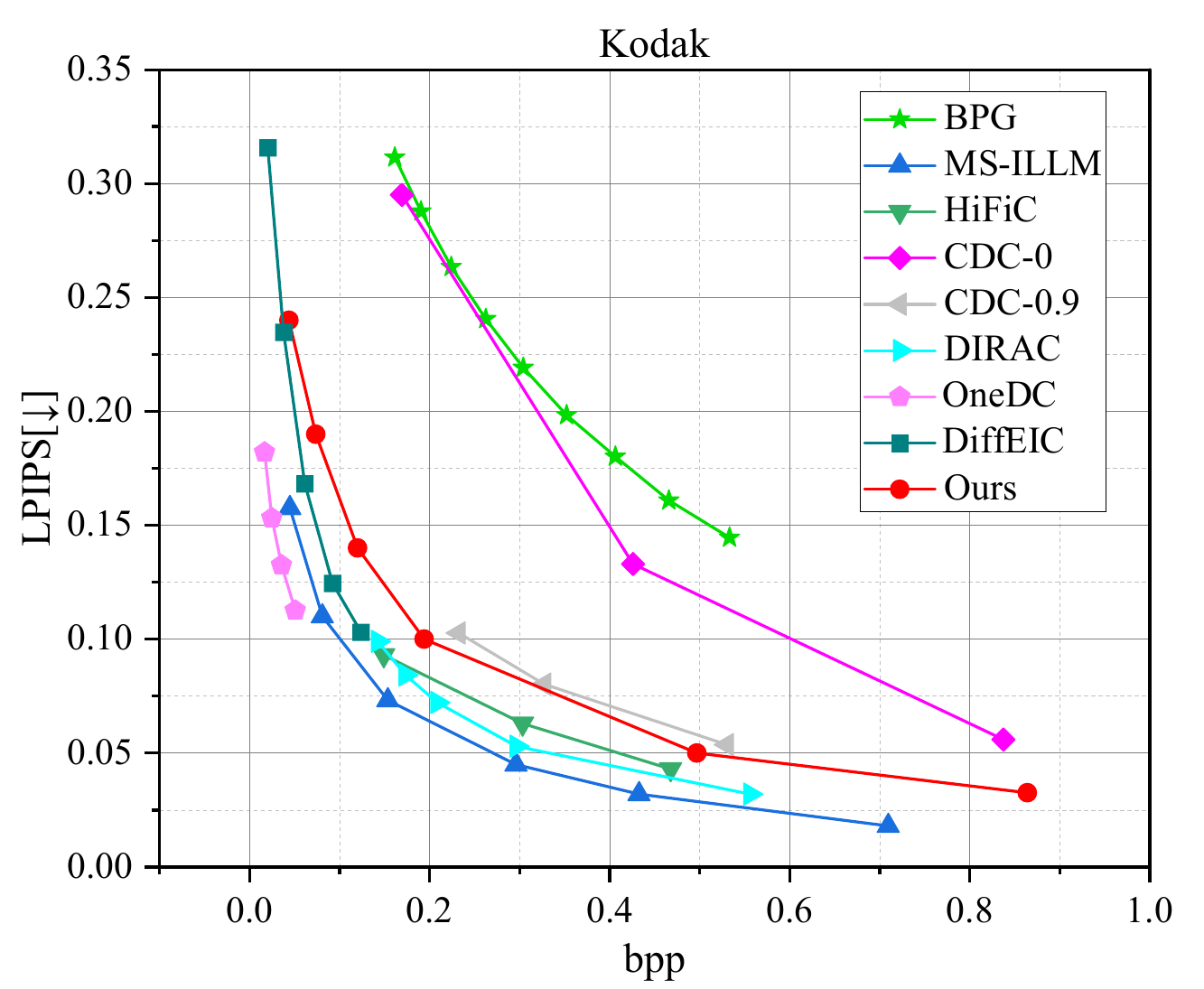}
    }
    \caption{Comparison of our method with existing methods in terms of rate-distortion $[\mathrm{bpp}\downarrow/\mathrm{PSNR}\uparrow]$ and rate-perception, including $[\mathrm{bpp}\downarrow/\mathrm{MS-SSIM}\uparrow]$ and $[\mathrm{bpp}\downarrow/\mathrm{LPIPS}\downarrow]$ for the Kodak dataset.}
    \label{fig5}
\end{figure*}

\begin{figure*}[h]
  \centering
    \includegraphics[width=1\textwidth]{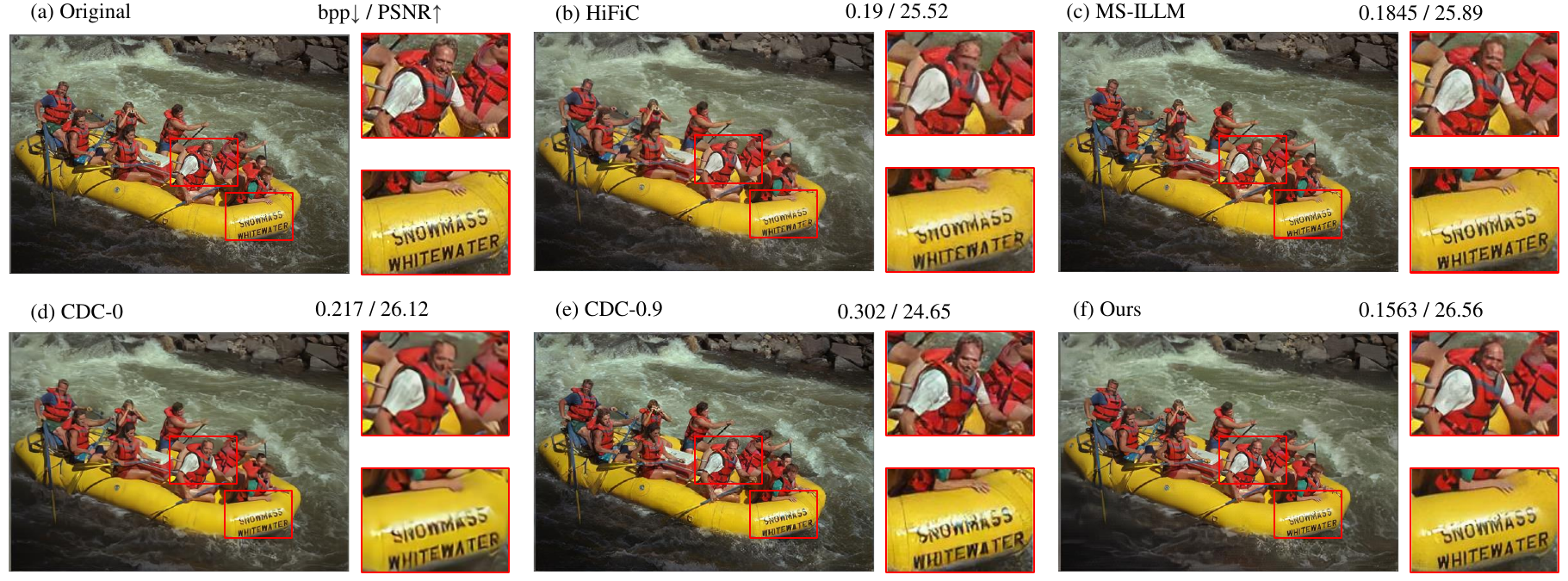}
    \caption{Visual comparison of our method against the existing methods on the reconstructed image $kodim14$ from Kodak dataset. Our method achieves better reconstruction quality with lower bitrates. In addition, our method displays fewer artifacts and more faithful reconstruction compared to other methods.}
    \label{fig5:main}
\end{figure*}

Specifically, a CLIP \cite{radford2021learning} perceptual loss $L_{CLIP}$ between $I_0^\theta$ and $I_T$ is calculated by the CLIP image encoder as:
\begin{equation}
  L_{CLIP}=\operatorname{E}[ ||D_{clip}(I_0^\theta) -D_{clip}(I_T) ||_2^2 ],
  \label{eq:13}
\end{equation}
where $D_{clip}(\cdot)$ denotes the features obtained by the CLIP image encoder, which contain the semantic information of the image. The mean value of the sampling process in the NC-Diffusion is adjusted by backpropagating this CLIP perceptual loss to the gradient of $I_0^\theta$ as: 
\begin{equation}
  I_{t-1}=I_t-(\bar{\alpha}_t-\bar{\alpha}_{t-1})\epsilon_\theta(I_t,t,I_T)-\lambda\nabla_{I_0^\theta}L_{CLIP},
  \label{eq:14}
\end{equation}
where $\nabla_{I_0^\theta}L_{CLIP}$ is a gradient of the CLIP perceptual loss and $\lambda$ represents the rate of the change made to the result. Applying a gradient of $L_{CLIP}$ to the mean value in the NC-Diffusion can force the prediction to retain the perceptual content of the initial reconstructed result during the iterative sampling process. 
\section{Experiments}

\subsection{Experimental Settings}

\begin{figure*}[htbp]
  \centering
    \includegraphics[width=1\textwidth]{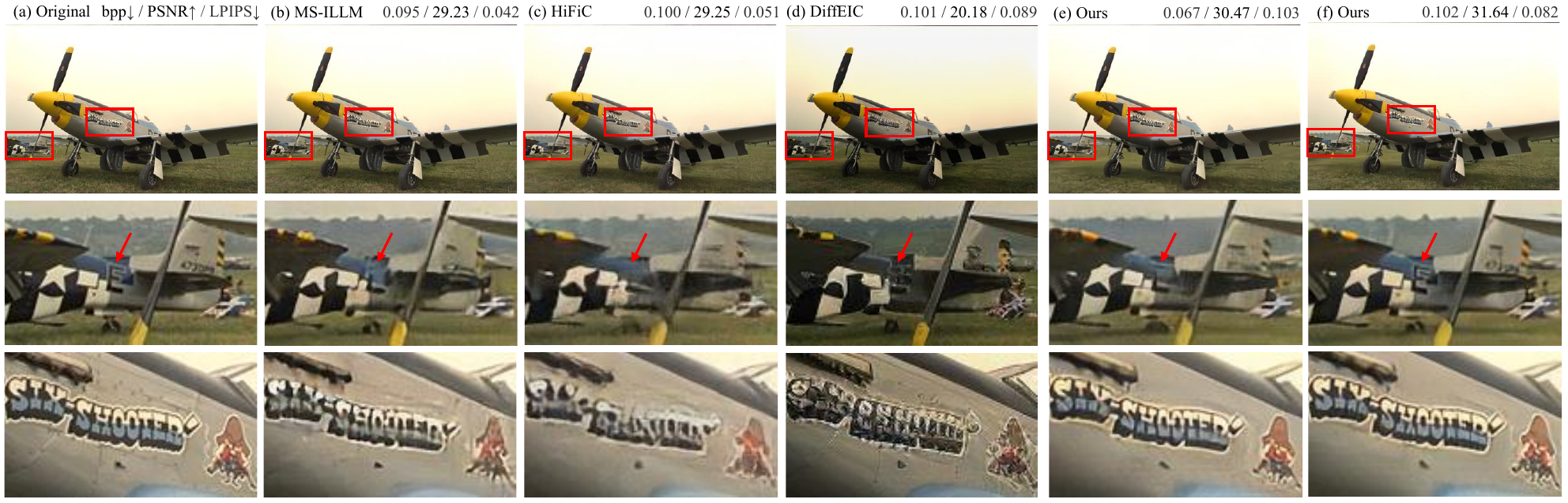}
    \caption{{Visual comparison of our method with the MS-ILLM \cite{muckley2023improving}, HiFiC \cite{mentzer2020high} and DiffEIC \cite{li2024towards} at a small bit-rate on the reconstructed image $kodim20$ from Kodak dataset.} Although MS-ILLM and HiFiC achieve better LPIPS scores, they deviate from the original image in terms of detailed texture such as the missing of letter ``E'' in the reconstructed images. By contrast, our method can better preserve the texture under similar or much smaller bitrates.}
    \label{lpips_ins}
\end{figure*}

\begin{figure*}[htbp]
    \centering
    \subfloat[Comparison on the CLIC2020 test dataset.\label{fig6:sub1}]{
        \includegraphics[width=0.4\textwidth]{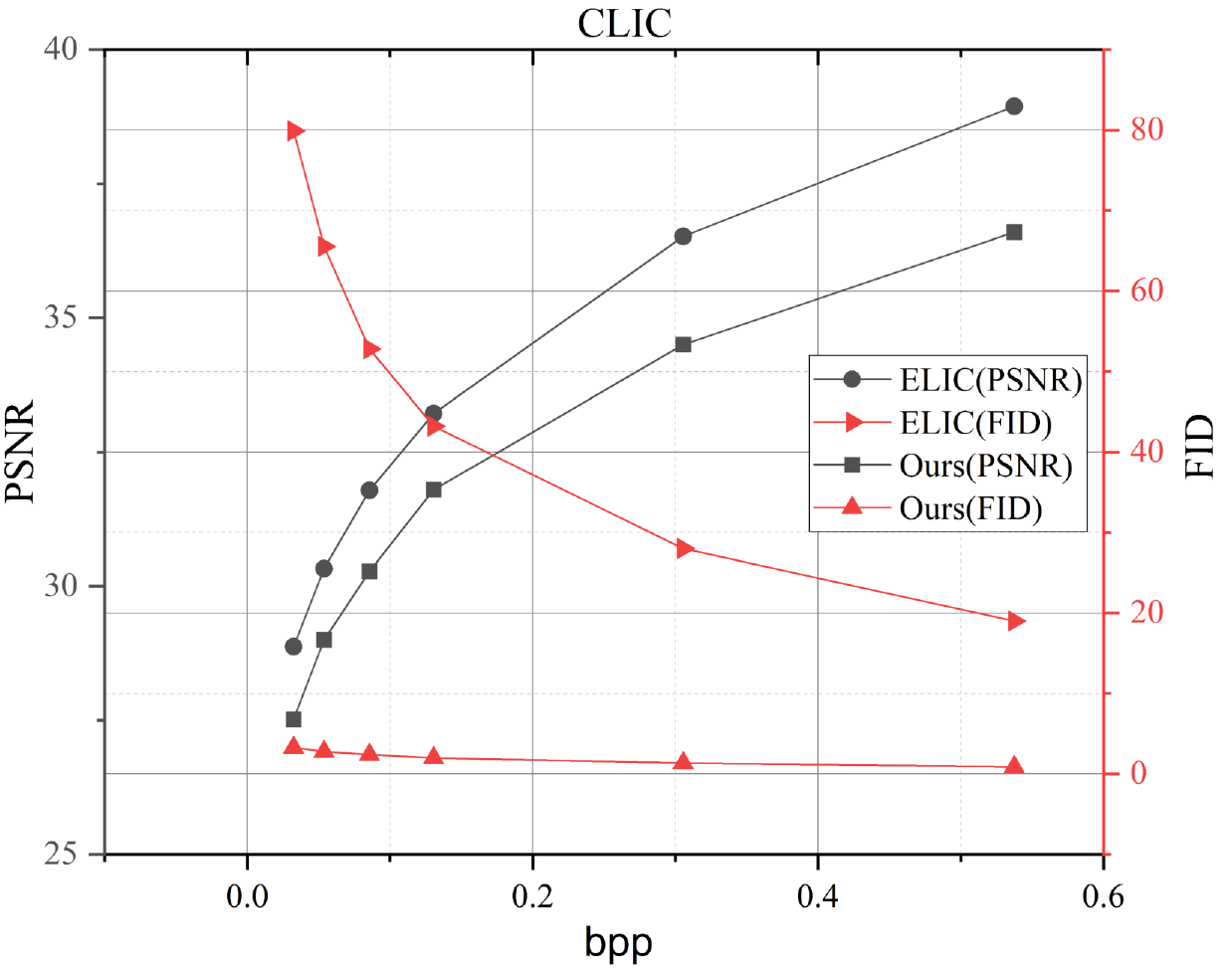}
    }
    \hspace{1.2cm}
    \subfloat[Comparison on the Kodak dataset.\label{fig6:sub2}]{
        \includegraphics[width=0.4\textwidth]{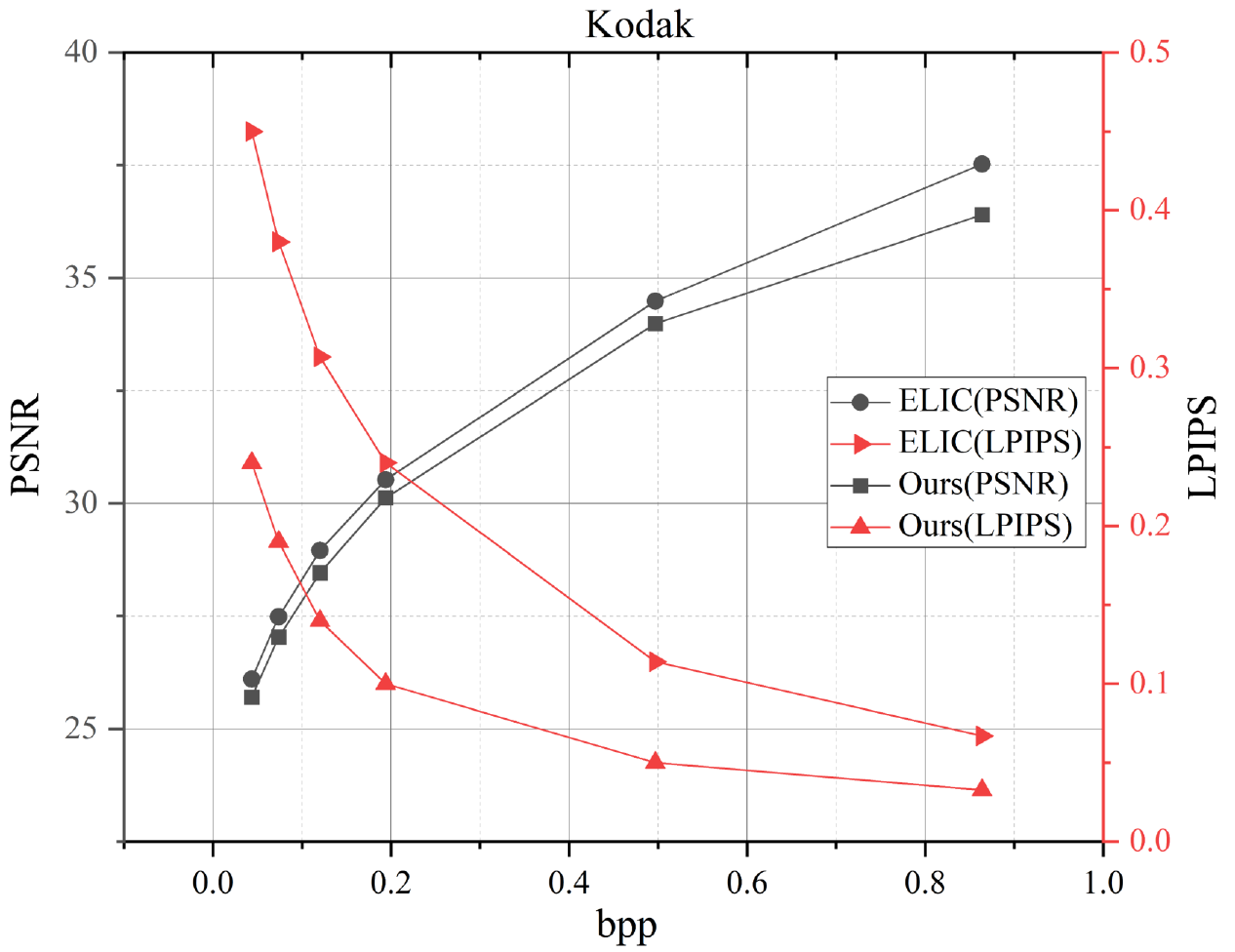}
    }
    \caption{The comparison of the distortion-perception tradeoff between our method and ELIC \cite{he2022elic}, on the CLIC2020 dataset (left) and Kodak dataset (right).}
    \label{fig6}
\end{figure*}

\textbf{Implementation details.} The Efficient learned image compression (ELIC) architecture \cite{he2022elic} is used as the neural compression backbone in our framework for initial compression. The pre-trained model is used to reduce the cost of acquiring paired datasets of initial compression results and ground-truth images used to train the NC-Diffusion. The training step T is set to 1000, and during inference, the sampler employs DDIM (Denoising Diffusion Implicit Models\cite{song2021denoising}) for accelerated sampling. In order to be able to perform inference process on images of arbitrary resolution without relying on the oversized GPU memory, overlapping approach is used to perform the inference process. The patch size is 256×256 and overlap size is 64 when inference process is performed on high resolution images. The backbone of the diffusion model employs a U-Net architecture (as shown in Fig. \ref{aff}) and other architectures such as DiT \cite{peebles2023scalable} can also be used. It contains a downsampling and corresponding upsampling process enhanced with skip connections to help reconstruct high-quality images with local details. Specifically, the downsampling path consists of four stages, where each stage comprises two residual blocks followed by a downsampling operation except the final stage. At the second stage, a self-attention layer is inserted after each residual block similarly as in \cite{saharia2022image,shang2024resdiff}. The upsampling path follows a similar structure, except that each stage contains three residual blocks to enhance its generation capability. The noise schedule in the NC-Diffusion uses a linear increasing strategy. The scale K of the wavelet transform is set to 4. The loss weight $\omega$ is set to 0.5 and the loss constant factor $\beta$ is set to 0.3 to balance the loss terms. The batch size is set to 1, the learning rate is set to 8e-5, and the number of training iterations is set to 80k using the Adam optimizer.

\textbf{Datasets.} The Flickr2W dataset \cite{liu2020unified} is used for training, which contains 20718 images. Two widely used benchmark test datasets, including the CLIC2020 test dataset \cite{clic2020dataset} and the Kodak dataset \cite{kodak}, are used for evaluation. The CLIC2020 test dataset consists of 428 images with a resolution of 2016 × 1512, and the Kodak dataset consists of 24 images with a resolution of 768 × 512.

\textbf{Evaluation metrics.} We evaluate our method and existing methods using BD-rate based on both distortion metrics and perceptual metrics. PSNR is used for distortion measurements, while MS-SSIM metric \cite{wang2003multiscale}, LPIPS metric \cite{zhang2018unreasonable} and FID metric \cite{heusel2017gans} are used for perceptual quality evaluation. In the same way as others \cite{agustsson2023multi,el2022image,hoogeboom2023high,mentzer2020high}, we also evaluate the FID metric at 256 × 256 resolution. For bitrates measurement, the compressed file size by performing the actual entropy coding process is used.


\subsection{Comparisons with State-of-the-art Methods}
Our method is evaluated with comparison to the state-of-the-art neural compression methods, including CDC \cite{yang2024lossy}, DIRAC \cite{ghouse2023residual}, HFD \cite{hoogeboom2023high}, HiFiC \cite{mentzer2020high}, MS-ILLM \cite{muckley2023improving}, DiffEIC \cite{li2024towards} and OneDC \cite{xue2025one}, which are all designed to focus on improving the perceptual quality of image compression results. CDC \cite{yang2024lossy}, DIRAC \cite{ghouse2023residual}, HFD \cite{hoogeboom2023high}, DiffEIC \cite{li2024towards} and OneDC \cite{xue2025one} are all diffusion-based image compression methods. Among them, DiffEIC \cite{li2024towards} and OneDC \cite{xue2025one} both employ the Stable Diffusion architecture and focus on extremely low bitrates image compression. CDC-0 and CDC-0.9 denote the diffusion-based image compression methods optimized entirely for distortion and jointly optimized for distortion and perception, respectively. In addition, the traditional image compression method BPG \cite{bellard2015bpg} is also used for evaluation.

The RD curves, including both rate-distortion trade-off and rate-perception trade-off on two benchmark datasets, are used for quantitative comparisons and illustrated in Fig. \ref{fig4} and Fig. \ref{fig5}. To algin with the results used in the existing methods, the different comparison metrics on the different datasets used by the existing methods are also adopted in our experiment as in \cite{hoogeboom2023high,ghouse2023residual}. Specifically, for the CLIC2020 test dataset, the FID metric is used. Note that the code of OneDC \cite{xue2025one} is not publicly available, and thus only the results reported in the paper, including PSNR, MS-SSIM and LPIPS, are compared without the results in terms of the FID metric. For the Kodak dataset, since it has only 24 images that cannot be used to measure the FID metric, the MS-SSIM metric is used to measure the perceptual quality. It can be found that our method outperforms the other state-of-the-art generative model compression methods both in terms of rate-distortion and rate-perception. For the compared methods, the results in terms of PSNR and perceptual quality are usually not consistent. Good perceptual quality usually means a relatively poor distortion metric. By contrary, our method can improve the perceptual quality while maintaining high PSNR performance. It is mostly because our method uses constrained noise in the diffusion process without introducing extra noise. 
\begin{figure}[t]
  \centering
    \includegraphics[width=0.4\textwidth]{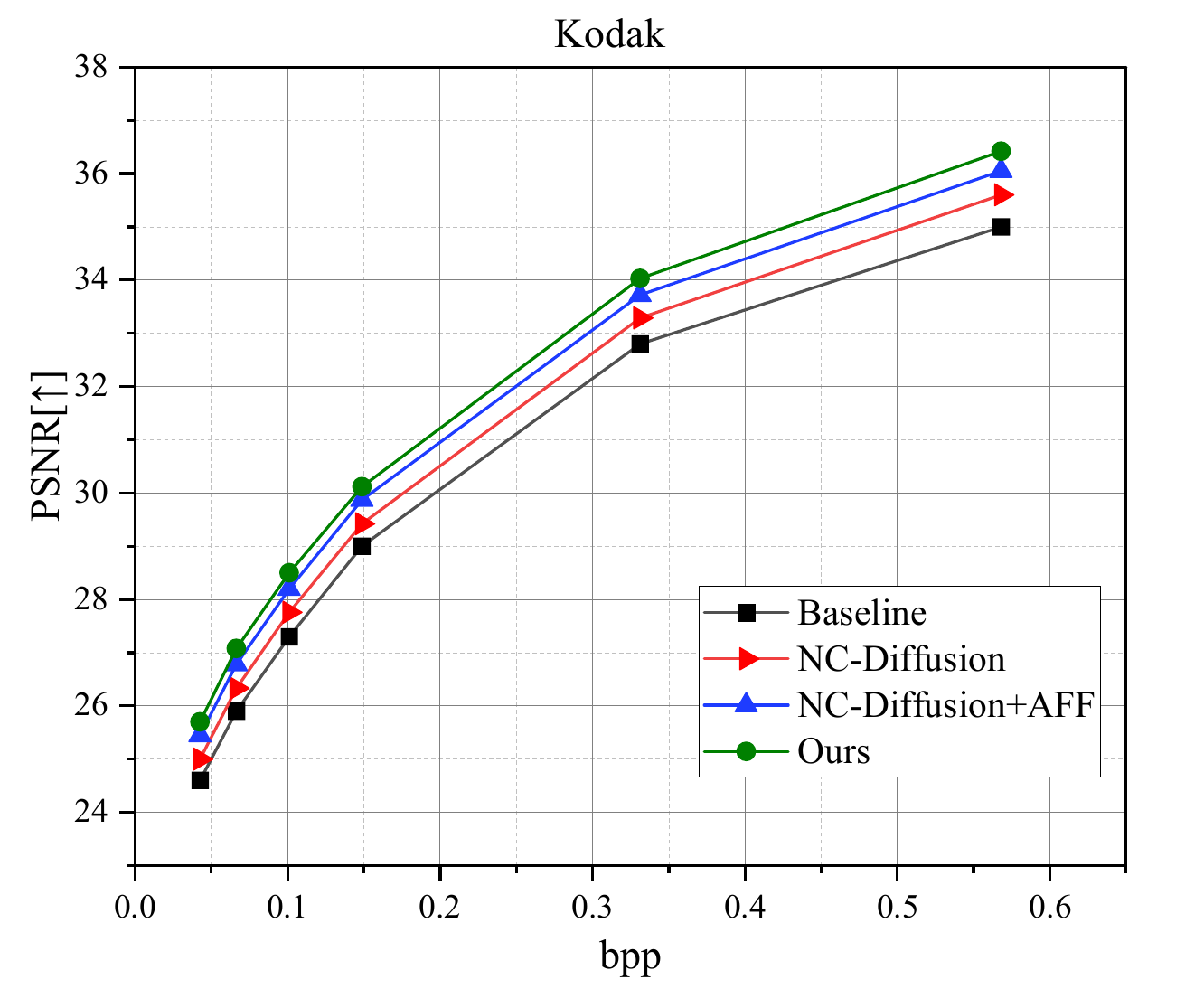}
    \caption{Comparison results of the ablation experiments on the Kodak dataset.}
    \label{abalation}
\end{figure}

Some example qualitative results of our method and the compared methods are visualized in Fig. \ref{fig5:main}. It can be seen that our method can produce results with better perceptual quality including more detailed textures, and better distortion metrics at lower bitrates. Moreover, for the Kodak dataset, the widely used LPIPS metric is also used for evaluation. However, this metric focuses more on the subjective quality of each reconstructed image and less on the faithful reconstruction of the ground-truth image. Therefore, the performance of our method is slightly worse than some of the generative methods such as MS-ILLM \cite{muckley2023improving} and HiFiC \cite{mentzer2020high} in terms of LPIPS metric. This is mostly because our method focuses on achieving better perceptual quality while maintaining the objective quality, a relatively faithful reconstruction.  To verify this, a visual comparison is further illustrated in Fig. \ref{lpips_ins}. It can be seen that although MS-ILLM \cite{muckley2023improving} and HiFiC \cite{mentzer2020high} achieve better LPIPS scores, they deviate from the original image in terms of detailed texture such as the missing of letter ``E'' in the reconstructed images. By contrast, our method can better preserve the texture under similar or much smaller bitrates.  Together with the better PSNR quality, it validates the effectiveness of our method in achieving a good tradeoff between distortion and perception. On the other hand, the FID metric can better evaluate the distribution alignment between the reconstructed image and the original image, and thus our method performs better on both PSNR and FID metrics as shown in Fig. \ref{fig4}.


\begin{table}[t]
  \centering
  {\caption{Comparison of the encoding and decoding speeds (seconds) per image on the Kodak dataset.}
  \label{tab1}
  \setlength{\tabcolsep}{5pt} 
  \renewcommand{\arraystretch}{1.3} 
  \begin{tabular}{l c c}
    \hline
    \textbf{Method} & \textbf{Encoding Speed (sec)} & \textbf{Decoding Speed (sec)} \\
    \hline
    ELIC & 0.042 & 0.063 \\
    HiFiC & 0.036 & 0.055 \\
    MS-ILLM & 0.038 & 0.057 \\
    CDC & 0.041 & 1.02 \\
    DiffEIC & 0.126 & 1.837 \\
    \textbf{NC-Diffusion (Ours)} & 0.042 & 0.102 \\
    \hline
  \end{tabular}}
\end{table}
\begin{figure*}[t]
  \centering
    \includegraphics[width=0.95\textwidth]{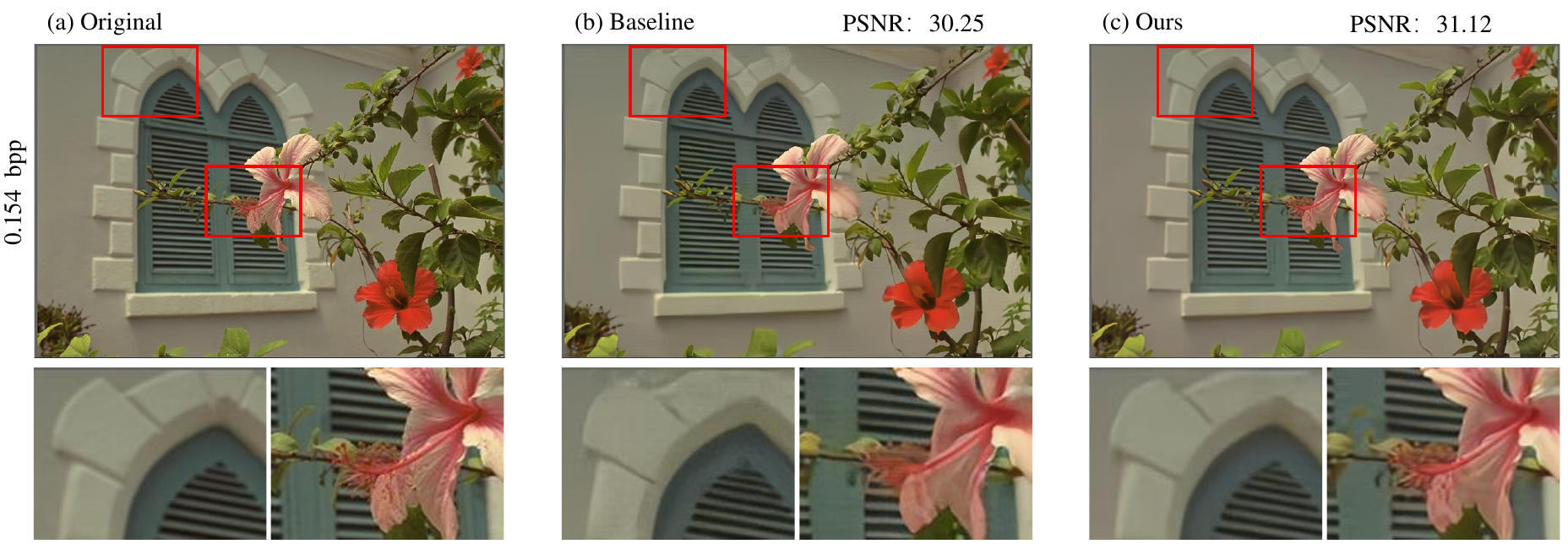}
    \caption{The visual comparison between our method and the baseline on the reconstructed image $kodim07$ from Kodak dataset.}
    \label{baseline_comp}
\end{figure*}

\begin{figure}[h]
  \centering
    \includegraphics[width=0.42\textwidth]{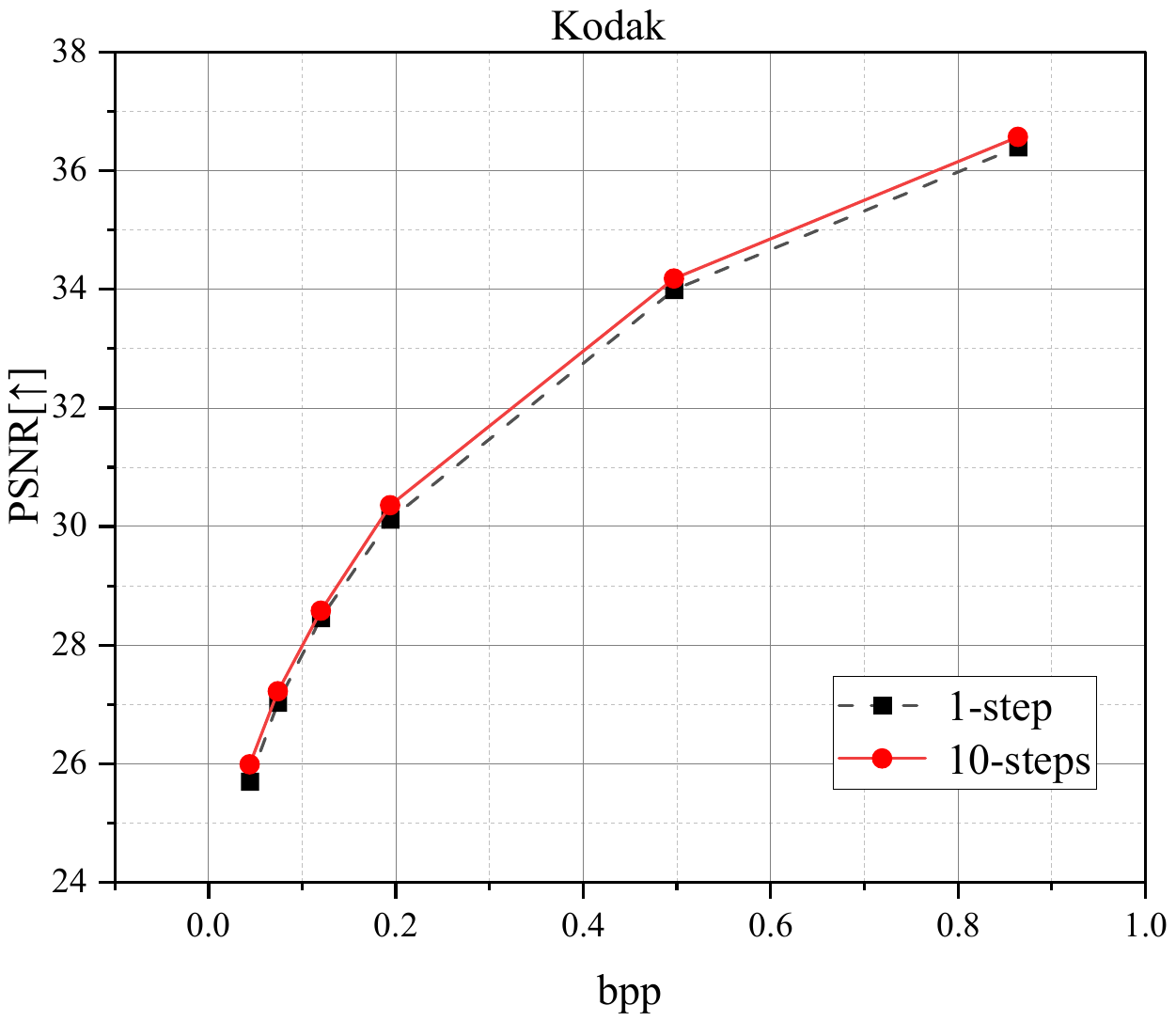}
    \caption{Quantitative comparison of the results obtained with different inference steps on the Kodak dataset.}
    \label{steps}
\end{figure}
To further validate that our method can achieve a better tradeoff between objective distortion and perception against the conventional distortion oriented compression methods, the performance of our method is further compared with the ELIC \cite{he2022elic} method. Fig. \ref{fig6} shows the performance comparison  in terms of distortion metric and perceptual metric. It can be observed that, compared to the ELIC \cite{he2022elic} method, our approach achieves a significant improvement in terms of the perceptual quality with only a small decrease in the objective quality. This is because the proposed NC-Diffusion guides the noise to redistribute in a more favorable manner for perception by re-sampling the quantization noise under different schedules and conducting iterative training at multiple noise scales. Furthermore, while a slight reduction in PSNR is observed, our method performs much better than other diffusion-based method while achieving greater perceptual improvement. This capability originates from the core mechanism of NC-Diffusion, which directly handles quantization noise instead of adding random Gaussian noise. Additionally, to comprehensively compare the overall performance of our method against ELIC \cite{he2022elic}, the BD-Rates are calculated based on the perceptual and PSNR metrics separately. On the Kodak dataset in terms of PSNR, ELIC \cite{he2022elic} provides 12.18\% BD-Rate saving compared to our method, while in terms of the LPIPS metric, our method achieves 70.49\% BD-Rate saving compared to ELIC \cite{he2022elic}. Considering both quality metrics, our method performs significantly better. In a word, our method outperforms existing generative compression methods \cite{yang2024lossy,ghouse2023residual,hoogeboom2023high, mentzer2020high,muckley2023improving,li2024towards,xue2025one} in both distortion and perceptual metrics and can achieve a good tradeoff between objective distortion and perception.

The complexity of the proposed NC-Diffusion, in terms of the encoding and decoding speed, is also evaluated in comparison with the existing methods on the Kodak dataset. The results are shown in Table \ref{tab1}. Compared to other diffusion-based approaches CDC \cite{yang2024lossy} and DiffEIC \cite{li2024towards}, the proposed method achieves significantly faster decoding. This is because the inference process of the NC-Diffusion starts from an initial reconstructed image rather than from random Gaussian noise, enabling it to produce high-quality results in just a single denoising step. In contrast, CDC \cite{yang2024lossy} and DiffEIC\cite{li2024towards} starts interference from random Gaussian noise and converge slower.

\begin{table}[t]
  \caption{Ablation study over different model components in terms of BD-Rate saving on the Kodak dataset.}
  \label{tab2}
  \centering
  \small
  \renewcommand{\arraystretch}{1.3} 
  \begin{tabular}{l c}
    \hline
    \textbf{Model Component} & \textbf{BD-Rate (\%)} \\
    \hline
    Baseline & 0 \\
    NC-Diffusion & -12.04 \\
    NC-Diffusion + AFF & -22.19 \\
    NC-Diffusion + AFF + $L_{high}$ & -28.01 \\
    \hline
  \end{tabular}
\end{table}

\subsection{Ablation Study}

In this section, we perform an ablation study to explore the effectiveness of each component in the proposed method. The Kodak dataset is used in the experiments. 

\begin{figure*}[t]
  \centering
    \includegraphics[width=0.95\textwidth]{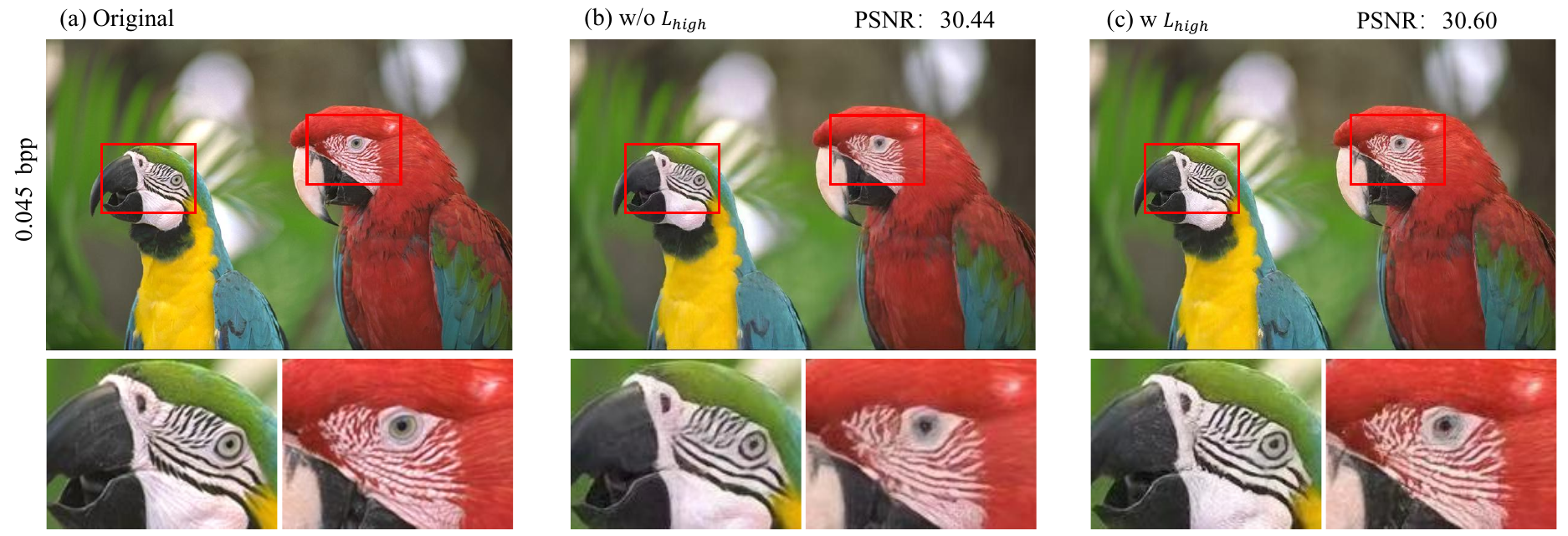}
    \caption{Visual comparison between the method without the $L_{high}$ loss term and the method with the $L_{high}$ loss term ($\beta$ is set to 0.3) on the reconstructed image $kodim23$ from the Kodak dataset.}
    \label{lhigh}
\end{figure*}

\textbf{Evaluation on the NC-Diffusion.} In order to verify the effectiveness of the proposed NC-Diffusion, the performance is compared with the general diffusion strategy as baseline, where the diffusion inference process is performed from a random Gaussian noise. The results, in terms of BD-Rate (\%), are shown in Table \ref{tab2}. Here the PSNR metric is used for BD-Rate calculation. It can be seen that the proposed NC-Diffusion effectively improves the performance, achieving over 12\% bitrates saving.  The RD curves comparison on different modules are also illustrated as shown in Fig. \ref{abalation}, demonstrating better performance at all rate points. The visual comparison between our method and the baseline is shown in Fig. \ref{baseline_comp}. It can be seen that our method can achieve better texture than the baseline method. This verifies that the constrained noise added to the diffusion and obtained from the quantization noise keeps more information of the texture details while avoiding introducing extra randomness into the reconstruction result, compared to the random Gaussian noise.

\textbf{Evaluation on the inference steps.} Although our method yields higher performance with just one inference step, we find that the distortion performance can be further improved by increasing the inference step. The RD curves of the PSNR metric at one inference step and at ten inference steps are compared to verify the effect of different inference steps. The results are shown in Fig. \ref{steps}. It can be seen that ten inference steps can further improve the performance. 

\begin{table}[t]
  \caption{Ablation study over different $\beta$ values on $L_{high}$ in terms of BD-Rate saving on the Kodak dataset.}
  \label{tab3}
  \centering
  \small
  \renewcommand{\arraystretch}{1.3} 
  \begin{tabular}{@{\hskip 0.4cm}p{3.13cm}  c@{\hskip 0.4cm}}
    \hline
    \textbf{Loss Function} & \textbf{BD-Rate (\%)} \\
    \hline
    $L_{diff}$ & 0 \\
    $L_{diff}+ 0.3 L_{high}$ & -7.73 \\
    $L_{diff}+ 0.6 L_{high}$ & -7.17 \\
    $L_{diff}+ 0.9 L_{high}$ & -4.66 \\
    \hline
  \end{tabular}
\end{table}

\textbf{Evaluation on the AFF module and the high-frequency detail preservation loss.} The effects of the high-frequency filtering and high-frequency loss on the diffusion network are further evaluated. The results are also shown in Table \ref{tab2}. It can be seen that both the proposed AFF and the high-frequency loss improve the performance. They encourage the NC-Diffusion to focus on learning high-frequency information, and enhance its ability to learn the detailed textures. {To further investigate the impact of different $\beta$ values on the $L_{high}$ loss term, an ablation study is conducted on the Kodak dataset. The results are shown in Table \ref{tab3}. 
It can be observed that as the weight of the $L_{high}$ loss term gradually increases, the performance first improves and then declines. This is because a large weight of the $L_{high}$ loss term may cause the model to focus more on the high frequency information, leading to a degradation in terms of the PSNR calculated on all frequencies. Additionally, a visual comparison of the results without and with the $L_{high}$ loss term is presented in Fig. \ref{lhigh}, demonstrating that the model incorporating the $L_{high}$ loss term can better restore textural details.}

\begin{figure}[t]
  \centering
    \includegraphics[width=0.46\textwidth]{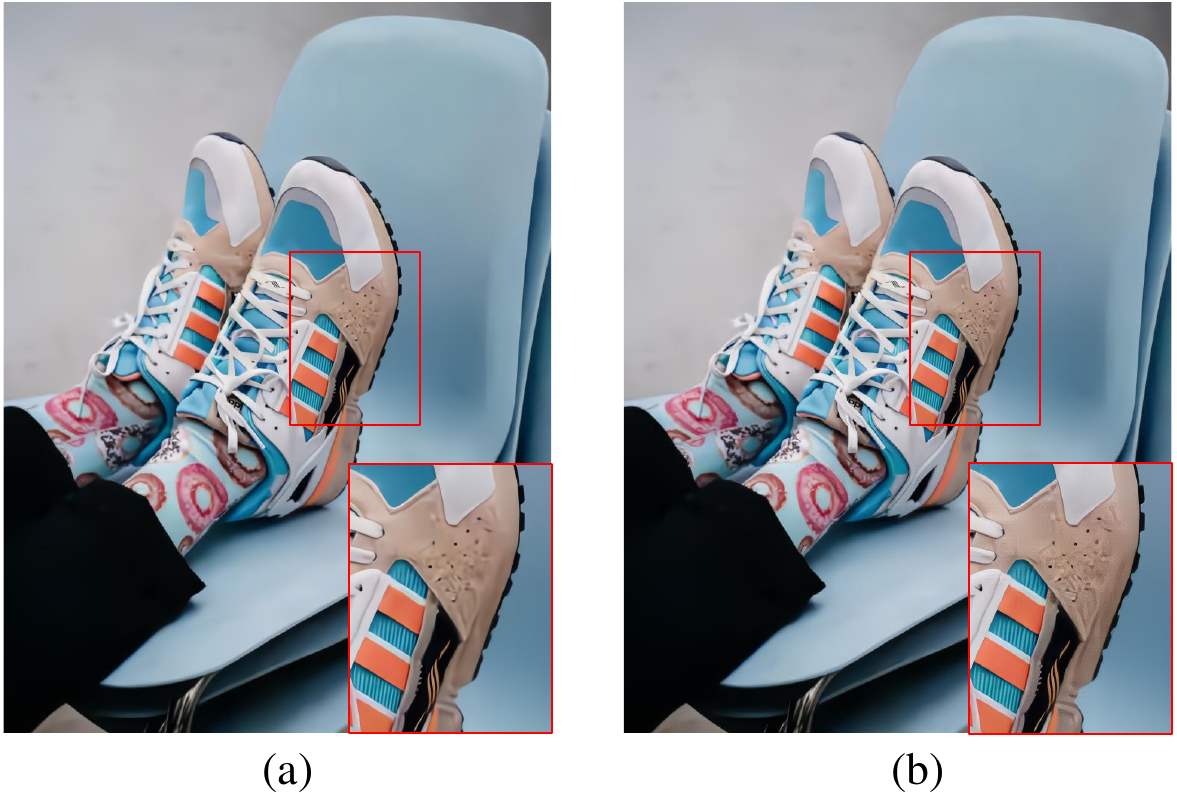}
    \caption{Visual comparison of the results obtained without (left) and with (right) sample-guided perceptual enhancement. It can be observed that the left image exhibits blurring artifacts in the shoe's ventilation holes area, while the right image demonstrates better perceptual quality.}
    \label{fig8:main}
\end{figure}

\textbf{Evaluation on the proposed zero-shot sample-guided enhancement.} The experimental results with and without this method are visualized for comparison, as shown in Fig. \ref{fig8:main}. The results are all generated by ten inference steps to better illustrate its effect since it is only used in test. From the results, it can be seen that the sample-guided method can help generate results with higher fidelity. This validates that the zero-shot sample-guided enhancement can trade-off distortion and fidelity performance in order to achieve perceptually satisfactory results.

\section{Conclusion}
In this paper, we propose a NC-Diffusion framework for high fidelity image compression. It solves the problem of noise mismatch when using diffusion for compression, by equivalating the quantization resulted noise in the learned image compression to added noise in the diffusion. A noise constrained diffusion process from the ground-truth image to the initial compression result is constructed to enable direct inference from the initial compression result without adding extra noise. Moreover, an adaptive frequency-domain filtering module combining high-frequency loss is developed to enhance the ability of the diffusion network by learning high-frequency information. Finally, a sample-guided enhancement method is designed to further improve the fidelity. Extensive experiments demonstrate that the proposed NC-Diffusion achieves better performance than the existing methods, verifying its effectiveness.

\bibliographystyle{IEEEtran}
\bibliography{main}

\end{document}